\documentclass{aa}  

\usepackage{graphicx}
\usepackage{txfonts}
\usepackage{xcolor}
\usepackage[dvipsnames]{xcolor}

\usepackage{lipsum}
\usepackage{subcaption}         
\usepackage{lscape}           
\usepackage{placeins}           
\usepackage[normalem]{ulem}
\usepackage{hyperref}
\usepackage{array}
\definecolor{cobalt}{rgb}{0.06, 0.2, 0.65}
\definecolor{violet}{rgb}{0.349, 0.149, 0.576}
\definecolor{khaki}{rgb}{0.84, 0.42, 0.01}
\hypersetup{
  colorlinks,
  citecolor=cobalt,
  linkcolor=[rgb]{0.8, 0.2, 1.0},
  urlcolor=[rgb]{0.8, 0.2, 1.0},
} 

\usepackage{environ,censor,calc}
\NewEnviron{mysout}
 {\soutrefunexp{\BODY}}
\newlength\nextcharwidth
\makeatletter
\renewcommand\@cenword[1]{
  \setlength{\nextcharwidth}{\widthof{#1}}
  \censorrule{\nextcharwidth}
  \kern -\nextcharwidth
  #1}
\makeatother
\newcommand\soutref[1]{\censorruledepth=.55ex\blackout{#1}}
\newcommand\soutrefunexp[1]{\expandafter\soutref\expandafter{#1}}

\newcommand{\ATLAS}{\object{3I/ATLAS}} 
\newcommand{\ObjSolarA}{\object{HD\,150469}}

\newcommand{\oi}{O\,{\sc i}}

\newcommand{\cni}{$^{12}\text{C}^{14}\text{N}$}
\newcommand{\cnii}{$^{13}\text{C}^{14}\text{N}$}
\newcommand{\cniii}{$^{12}\text{C}^{15}\text{N}$}
\newcommand{\hiio}{$\text{H}_{2}\text{O}$}
\newcommand{\coii}{$\text{C}\text{O}_{2}$}
\newcommand{\ooned}{O(\ensuremath{^1\mathrm{D}})}
\newcommand{\oones}{O(\ensuremath{^1\mathrm{S}})}

\usepackage{natbib}
\bibpunct{(}{)}{;}{a}{}{,} 

\begin{document}

   \title{Very Large Telescope observations of interstellar comet \ATLAS}

   \subtitle{III: High-resolution monitoring of CN and forbidden oxygen emission
across the perihelion passage with ESPRESSO}

   \author{Baltasar Luco\inst{1}\corrauth{baltasarluco@uc.cl}     
        \and Thomas H. Puzia\inst{1}\email{tpuzia@astro.puc.cl}
        \and Rohan Rahatgaonkar\inst{1}\email{rrohan@uc.cl} 
        \and Juan Pablo Carvajal\inst{1}\email{jcarvajal000@gmail.com}
        \and Prasanta K. Nayak\inst{1}\email{rrohan@uc.cl} 
        }

   \institute{Institute of Astrophysics, Pontificia Universidad Cat\'olica de Chile, Av.~Vicu\~na Mackenna 4860, 7820436 Macul, Santiago, Chile}

   \date{Received 29 April 2026 / Accepted 14 July 2026}

  \abstract
    {\ATLAS\ is only the third known interstellar object to traverse the Solar System, and the first object for which both branches of the perihelion passage can be monitored at high spectral resolution.}
    {We characterize in \ATLAS\ the heliocentric evolution of the CN production rate and the forbidden [O\,\textsc{i}] emission, describe the broader emission line inventory, and release the analysis tools and reduced data products to the community.}
    {We obtained high-resolution ($R\!\simeq\!140\,000$) spectroscopy of \ATLAS\ on 23 nights with VLT/ESPRESSO ($r_{\rm h}\!=\!1.67\!-\!2.45$\,au). The CN production rates were derived by fitting the $B^2\Sigma^+\!-\!X^2\Sigma^+$ band using {CometSpec}, a publicly released Python package for flexible fluorescence modeling. The three forbidden [O\,\textsc{i}] lines were modeled to derive the green-to-red (G/R) flux ratio and a \coii/\hiio\ proxy. The two quantities were fitted with symmetric and piecewise-asymmetric power laws.}
    {We detected CN in 18 of 22 useful nights. We fit $\text{Q}_{\rm CN}=a\,(r_{\rm h}/1\text{ au})^{\,b}$ to the production rate versus heliocentric distance relation. The pre-perihelion CN power-law index is $b_{\rm pre}\!=\!-4.62^{+1.25}_{-1.22}$, which is within the range observed for Solar System comets. The CN residuals about the smooth model show substantial scatter that we attribute to a combination of systematics and possible physical drivers that the current cadence cannot disentangle. The G/R ratio decreases from $\sim\!0.42\!\pm\!0.14$ at $r_{\rm h}\!\simeq\!2.4$\,au to $\sim\!0.114\!\pm\!0.004$ near perihelion. The asymmetric piecewise oxygen fit is strongly preferred over the symmetric fit, consistent with thermal inertia models, although a denser post-perihelion sampling is needed to confirm this. The \coii/\hiio\ proxy spans $\sim\!0.06\!-\!1.06$ and is comparable to the values reported by Subaru/HDS and SPHEREx and to those measured for 2I/Borisov. We additionally identify Ni\,\textsc{i} and Fe\,\textsc{i} and report post-perihelion production rates for $\rm C_2$ and CH.}
    {Our high-resolution monitoring of \ATLAS\ reveals a \coii-rich coma that becomes progressively \hiio-dominated near perihelion, with a G/R asymmetry consistent with delayed \hiio\ sublimation predicted by thermal inertia models. Together with the public release of \textsc{CometSpec} and the reduced data, these results provide an empirical reference and a methodological framework for the next generation of interstellar object monitoring campaigns.}


\titlerunning{VLT observations of interstellar comet \ATLAS\ III}
\authorrunning{Luco, B., et al.}
   \maketitle
    \nolinenumbers

\section{Introduction}\label{sec:intro}

Comets offer a unique window into the chemistry of cosmic ices and the radiative and collisional processes operating in tenuous gas. As a comet approaches the Sun, the differential sublimation of the parent compound and the subsequent photodissociation of the compounds defines the chemical pathways traced by the species we observe in the coma. Comprehensive monitoring of different comet types is the foundation for building a self-consistent and complete dataset that enables a thorough study of their chemistry and morphology. In this work, we focus on two complementary diagnostics accessible to high-resolution optical spectroscopy: the CN violet band, and the forbidden [O\,\textsc{i}] lines.

The CN $B^2\Sigma^+$--$X^2\Sigma^+$ violet band is one of the most prominent features in cometary spectra and was among the first molecules detected in comets, remaining detectable beyond $r_{\rm h}\!\simeq\!3$\,au owing to its high fluorescence efficiency. CN is produced primarily by photodissociation of HCN \citep{Combi80}, although in some cases, additional production channels must be invoked to explain the observed CN abundance \citep[e.g.,][]{Fray05}. When the spectral resolution and signal-to-noise ratio are high enough, the isotopic ratios $^{12}$C/$^{13}$C and $^{14}$N/$^{15}$N can be retrieved from individual rotational lines of the violet band \citep{Manfroid09, Bockelee-Morvan15, Moulane23}. The carbon ratio is set primarily by stellar nucleosynthesis and Galactic chemical evolution, while the nitrogen ratio records the UV environment and radial location within the natal protoplanetary disk where the nitrogen-bearing ices were assembled. Together, these isotopic ratios trace the Galactic chemical evolution and the local formation environment of the parent body (for comprehensive reviews, we refer to \citet{Furi15} and \citet{Romano22}).

After hydrogen, oxygen is the second most abundant element in comets. It is mainly present in the form of CO, \coii, and \hiio. In the optical, oxygen is observable through the forbidden lines arising from the metastable \ooned\ and \oones\ excited states. The 5578\,\AA\ green line is sensitive to the combined contribution of CO, \coii, and \hiio, whereas the red doublet at 6302 and 6365\,\AA\ is driven primarily by \hiio\ photodissociation \citep{Festou81, Fink84, Combi91}. Because no electronic transitions of \hiio\ and \coii\ are accessible at optical wavelengths and their rotational and vibrational bands fall in spectral regions affected by strong telluric absorption, the two molecules are difficult to detect directly from the ground. As a consequence, the green-to-red flux ratio (G/R) of the forbidden oxygen lines has been established as an indirect proxy for the \coii/\hiio\ ratio in the coma \citep[e.g.,][]{Decock13, McKay15}, which in turn evolves with heliocentric distance owing to the different sublimation temperatures of these molecules ($152$~K for \hiio\ and $72$~K for \coii). A practical complication when measuring the cometary [O\,\textsc{i}] lines is contamination by telluric oxygen emission. At a sufficiently high spectral resolution, however, the cometary and telluric components can be clearly separated through the geocentric velocity of the comet \citep[e.g.,][]{Moulane23}. The 5578\,\AA\ line can additionally be blended with C$_2$ features, and this degeneracy can likewise be resolved at high resolving power.

Many other molecules have been identified in comets, each providing key diagnostic information. Carbon-chain species, for instance, can emit strongly at visible wavelengths. 
Although their chemistry is still under study, they are useful probes of the coma as daughter species of parents such as $\rm C_2H_2$ \citep{Pierce21}. Comets have also been classified as carbon-chain depleted or rich \citep{AHearn95}. This distinction is thought to reflect the primordial composition of the region in which they formed. Other relevant species include CH, part of the hydrogen-bearing carbon chemistry, and $\rm NH_2$, an important nitrogen-bearing molecule whose ortho-to-para ratio can serve as a temperature tracer \citep{Kawakita04}. More recently, Fe and Ni have been detected in comets even at large heliocentric distances \citep{Manfroid21}, opening the study of Fe- and Ni-carbonyl chemistry. Many further species and applications remain to be explored. In all cases, a robust characterization requires long-term monitoring, a sufficiently high signal-to-noise ratio, and if possible, a spectral resolution that is high enough to separate the lines of the same or different species, so that the emission is captured at high fidelity.

Interstellar objects (ISOs) traversing the Solar System have long been predicted as a natural consequence of planetesimal ejection during planetary system formation \citep{McGlynn89, Jewitt03}, and they offer a unique opportunity to study the chemistry of their parent system. The discovery of 1I/`Oumuamua on October 19, 2017 \citep{Meech17}, confirmed this prediction, but the first ISO proved difficult to characterize: with a strongly hyperbolic orbit ($e \approx 1.2$) and an unusual elongated morphology, 1I displayed no detectable coma or gas emission despite intensive observational campaigns, and only upper limits could be placed on the production rates of CN, C$_2$, and C$_3$ \citep{Ye17}. Non-gravitational acceleration along its trajectory hinted at outgassing below detection thresholds \citep{Micheli18}, but the absence of observable volatiles, combined with the post-perihelion timing of its discovery, precluded any compositional characterization of its ices \citep{Fitzsimmons18, OumuamuaISSIteam2019}.

The second confirmed ISO, 2I/Borisov, discovered on August 30, 2019 \citep{Borisov2019MPEC}, provided the first opportunity for a detailed spectroscopic study of interstellar cometary material. 2I/Borisov developed an active dust coma and was discovered well before its perihelion at $q\!=\!2.0$\,au, enabling a comprehensive monitoring campaign. Optical spectroscopy detected CN at $r_{\rm h}\!=\!2.7$\,au \citep{Fitzsimmons19}, followed by C$_2$ \citep{Lin20}, [O\,\textsc{i}] \citep{McKay20}, OH \citep{Xing20}, NH$_2$ \citep{Bannister20}, and exceptionally high CO abundances \citep{Bodewits20, Cordiner20}. The detection of atomic nickel and iron vapor in 2I/Borisov \citep{Opitom21} expanded the inventory of detected species, which was unusual at the time, but was rapidly recognized as ubiquitous in Solar System comets \citep{Manfroid21, Bromley21}. Overall, 2I/Borisov appeared to be broadly similar to Solar System comets in its gas and dust properties, although it showed distinctive compositional features such as a high CO abundance. Neither 2I/Borisov nor 1I/`Oumuamua, however, benefited from a multi-night monitoring campaign spanning the pre- and post-perihelion phases, and 1I/`Oumuamua lacked detectable molecular emission altogether.

3I/ATLAS was discovered on July 1, 2025, by the Asteroid Terrestrial-impact Last Alert System \citep[ATLAS;][]{Tonry18} at a heliocentric distance of $\sim\!4.5$\,au on an incoming strongly hyperbolic trajectory with an eccentricity $e\!\approx\!6.14$ \citep{Denneau25, Seligman25, Hopkins25}. The early discovery, several months before perihelion on October 29, 2025 ($q\!=\!1.356$\,au), enabled an unprecedented multi-facility campaign spanning ground- and space-based observatories. Initial spectroscopy revealed a red, dust-dominated coma reminiscent of D-type asteroids and trans-Neptunian objects \citep{Opitom25, Puzia25, Seligman25}, with subsequent detections of CN and Ni\,\textsc{i} emission evolving rapidly as 3I/ATLAS approached the Sun \citep{Rahatgaonkar25}. Near-infrared spectroscopy with the James Webb Space Telescope (JWST) revealed a coma unusually rich in CO$_2$ and CO relative to water \citep{Cordiner25, Belyakov26}. The Spectro-Photometer for the History of the Universe, Epoch of Reionization and Ices Explorer (SPHEREx), however, placed the CO$_2$ and CO abundances of 3I/ATLAS as elevated but consistent with those of long-period Solar System comets \citep{Lisse26}. During the pre-perihelion phase, the absence of detectable C$_2$ and C$_3$ placed 3I/ATLAS among the most carbon-chain-depleted comets known \citep{Salazar25, Rahatgaonkar25}. This classification was subsequently revised following post-perihelion detections of carbon-chain species, bringing 3I/ATLAS back into broad agreement with Solar System comets \citep{Kawakita26}. The detection of atomic nickel without accompanying iron \citep{Rahatgaonkar25} pointed to a carbonyl parent species as the driver of Fe and Ni chemistry in comets, and \citet{Hutsemekers26} further showed that this mechanism causes the Ni/Fe ratio to converge toward the Solar System trend at decreasing heliocentric distances. Together, these findings highlight the strong time dependence of cometary spectroscopic signatures and underscore the importance of monitoring the volatile evolution of comets, especially ISOs. 3I/ATLAS thus provides a foundation for and a benchmark against all future ISO studies.

In this paper, we present high-resolution spectroscopy ($R\!\simeq\!140\,000$) from the Very Large Telescope (VLT) Echelle SPectrograph for Rocky Exoplanets and Stable Spectroscopic Observations (ESPRESSO) of the interstellar comet 3I/ATLAS, obtained at heliocentric distances $r_{\rm h}\!=\!1.67\!-\!2.45$\,au across the pre- and post-perihelion branches of its orbit. We report
\begin{enumerate}
    \item the heliocentric evolution of the CN production rate $\rm Q_{\rm CN}$ at 22 epochs and also discuss the observed scatter in the context of systematics and physical processes that may drive it.
    \item the detection and modeling of the three forbidden [O\,\textsc{i}] lines at 5578, 6302, and 6365\,\AA\ with simultaneous two-component (telluric plus cometary) Voigt profiles, yielding the green-to-red ratio (G/R), an indirect \coii/\hiio\ proxy, and per-epoch radial velocities at the level of hundreds of m\,s$^{-1}$.
    \item the identification of emission features, including atomic nickel and iron lines and molecular bands of CH and C$_2$, with comparisons between pre- and post-perihelion stacks and between nucleus-centered and offset coma sight lines.
    \item the public release of \textsc{CometSpec}, a \textsc{Python} package for cometary fluorescence modeling and Markov chain Monte Carlo (MCMC) fitting, incorporating rotational-collisional coupling, MCMC parameter estimation, isotope modeling, and support for arbitrary user-supplied line lists for other molecular species.
\end{enumerate}

The paper is organized as follows. Section~\ref{sec:obs} describes the observations and observing strategy. Section~\ref{sec:red} details the data reduction, including sky subtraction, telluric line masking, stacking, and continuum modeling. Section~\ref{sec:analysis} presents the CN fluorescence modeling and fitting, the derivation of production rates, the forbidden oxygen line analysis, the wavelength-calibration cross-check, and the broader emission line inventory. Section~\ref{sec:disc} discusses the heliocentric evolution of $\rm Q_{\rm CN}$, the G/R ratio, and the implied \coii/\hiio\ trend. Section~\ref{sec:conc} summarizes our conclusions and outlines future directions. Section~\ref{sec:opensource} provides the links to our data and software, and the appendix provides additional detail on \textsc{CometSpec} and a complementary analysis.

\section{Observations}\label{sec:obs}

We used the high-resolution spectrograph ESPRESSO at the ESO 8.2\,m VLT on Cerro Paranal in Chile \citep{Pepe21}. We observed the interstellar comet 3I at heliocentric distances $r_{\rm h}\! =\! 1.67\!-\!2.45$\,au as part of an ESO director’s discretionary time program (Prog. ID 115.29J6, PI: Puzia). The observations were carried out with the high-resolution setup, which provides a resolving power of $R\!\simeq\!140\,000$ in the wavelength range $380$–$788$\,nm. The main fiber (A; $1\arcsec$ diameter) was centered on the comet nucleus, while the second fiber (fiber~B) was used in SKYMODE. Fiber B is radially separated by $7\arcsec$ on the sky from the main fiber, and in the case of 3I, samples the coma of the comet, rather than blank sky. Because ESPRESSO can use any available VLT unit telescope (UT), the position angle of fiber~B on the sky depends on the UT that was used for a particular epoch. As there is no derotator, fiber~B rotates on the sky during the exposures around fiber~A. The physical radius range at the $r_{\rm h}$ distances of 3I, covered by fiber~A, spans $\sim\!653-932$\,km from the comet nucleus, and fiber~B is positioned at a projected nucleocentric distance of $\sim\!9\,140\!-\!13\,050$\,km. As fiber~B does not provide a pure sky background, separate sky observations were taken interleaved with the science exposures, ensuring that the sky frames were obtained as close in time as possible to the science observations. However, as the air mass of 3I increased through September, the number of sky observations was reduced to prioritize the acquisition of science flux, first to a single sky frame, and eventually, to no dedicated sky acquisition\footnote{We exclusively focus on emission features, not on the reflectance spectrum. At the spectral resolution of ESPRESSO, telluric emission lines can be identified and masked even on nights without dedicated sky exposures. The absence of sky frames therefore does not affect our analysis and results.}. To guarantee good tracking and fiber-A centering, we performed reacquisition of 3I in between observation blocks.

The journal of observations is provided in the Table~\ref{tab:observations}, where we also summarize some relevant comet parameters extracted from the Jet Propulsion Laboratory (JPL) Horizons System\footnote{\url{https://ssd.jpl.nasa.gov/horizons/}}. Due to technical issues, the night of September 21 was discarded from the analysis. In addition, we removed one exposure taken on September 9 due to star contamination and discarded another exposure from the December 6 run due to bad observing conditions.

\section{Data reduction}\label{sec:red}

We used the standard ESOREFLEX recipes \citep{Freudling13} to reduce the data, adopting the default parameters and the standard calibration frames. After the initial reduction, we used the telluric-corrected (H$_2$O only) and flux-calibrated output of the pipeline as our 3I fiber~A spectrum ($F^A_{\rm 3I}=F^A_{\rm total} - F^A_{\rm telluric}$), without the default sky-subtraction option, which normally is the flux in fiber~B ($F^B_{\rm total}$). The pipeline does not output the reduced flux from fiber~B ($F^B_{\rm 3I}$). We therefore recovered the 3I fiber B spectra using the default sky-subtraction option for fiber~A switched on ($F^A_{\rm total}-F^B_{\rm total}$) in the following way:
\begin{eqnarray*}
    F^B_{\rm 3I} &=& F^A_{\rm 3I} - (F^A_{\rm total}-F^B_{\rm total}).
\end{eqnarray*}

From these spectra, we first ensured a precise wavelength calibration. We performed a Gaussian (plus constant) Levenberg-Marquardt least-squares (LS) fit to the [\oi] $5578.8955$\,\AA\,sky line \citep[vacuum reference value from][]{Hanuschik03} and forced the solution to be centered at this value using a velocity correction. This procedure was only applied to fiber~A. Since fiber~B shares the wavelength solution of the main fiber, this and any subsequent wavelength correction were derived from fiber~A and then applied to fiber~B. Additionally, since this sky line in our observations tends to be stronger than the corresponding cometary line, and combined with the high resolving power and a good initial wavelength solution, the use of a single-Gaussian fit to this feature achieved the accuracy and precision within the range needed to compare the spectra taken on different nights. This was demonstrated a posteriori by a cross-check based on the three forbidden [\oi] lines, modeled with a two-Voigt fit after applying all wavelength corrections (see Sect.~\ref{sec:geocentric}).

At the resolution of ESPRESSO, the spectral flux density of the sky continuum emission is weak. For nights with dedicated sky exposures, the sky was modeled as a first-order polynomial using a variance-weighted linear least-squares fit with iterative sigma clipping (three iterations at $\sigma = 3$). For science exposures on these nights, sky subtraction was performed using the closest sky exposure in time. We modeled the continuum flux following \citet{Rahatgaonkar25} with two adaptations. First, we fixed the heliocentric velocity to the ephemeris value obtained from JPL Horizons. Second, we used the Kurucz solar irradiance \citep{Kurucz05}\footnote{\url{http://kurucz.harvard.edu/sun/IRRADIANCE2005/}} instead of a solar analog. We emphasize, however, that at our resolution, the lines are sufficiently resolved to model the sky and the reflectance continuum jointly and do not affect the measured line fluxes. We analyzed several stacked spectra: (i) per-night stacks, (ii) full pre-perihelion stacks, (iii) full post-perihelion stacks, (iv) the three latest pre-perihelion epoch stacks, and (v) the three earliest post-perihelion epoch stacks\footnote{These stacks were produced following the same procedure, with the difference that the continuum and Kurucz irradiance in each stack was computed as the weighted mean of the continuum and Kurucz irradiance of the individual nights stacks, all interpolated to the same wavelength grid.}.

Telluric absorption and sky emission lines do not significantly affect the emission lines we analyzed. Nevertheless, our final data products include spectra for which the identified sky emission lines were masked.

The spectra were shifted to the comet rest frame using the geocentric velocity $\dot{\Delta}$ (at the observatory) and then stacked on a per-night basis. For each night, all spectra were linearly interpolated onto the wavelength grid of the first comet exposure of that night (tests with alternative interpolation schemes produced only minimal changes). The stacking was performed with a variance weighted mean with a $5\sigma$ clipping of three iterations to exclude as many bad pixels as possible, and only wavelength points with at least two valid contributors were retained. The error of the stacked spectrum was obtained by Gaussian propagation of the individual uncertainties. Bad pixels are strongly down-weighted by the variance weighting, and although some may remain, they do not affect our analysis.

By applying this procedure, we obtained a final data product for each night, consisting of the per-night stacked flux and error spectrum in the wavelength restframe of the comet, a spectrum with the number of exposures contributing to each wavelength point, another flux and error spectrum with the number of contributors after masking telluric lines, the modeled continuum, the modeled reflectance, and the Kurucz solar irradiance spectrum shifted accordingly to the comet restframe. The products are publicly available\footnote{\url{https://zenodo.org/records/20966879}}. We emphasize that the reflectance values obtained with this method are meaningful only for nights on which a sky observation was obtained.

\section{Analysis}\label{sec:analysis}

\subsection{CN modeling}\label{sec:cn}
We developed a Python package for fluorescence modeling, \textsc{CometSpec}, which broadly follows the prescriptions by \cite{Manfroid09} and \cite{Bromley24} and is publicly available on GitHub\footnote{\url{https://github.com/baltasarluco/CometSpec}}. The key ingredients required to build a fluorescence model are (i) the solar irradiance spectrum that induces radiative transitions between molecular states, and (ii) the set of spectral lines to be considered in the model.

For the solar irradiance, we used the high-resolution spectrum from the latest Kurucz solar irradiance compilation. This spectrum has a wavelength sampling of approximately 0.005~nm and covers the wavelength range $\sim\!300\!-\!1000$~nm. The irradiance was scaled to the heliocentric distance of the comet and Doppler-shifted according to its heliocentric velocity for each observing night, in order to account for the Swings effect. Lines beyond the provided solar irradiance are excluded by \textsc{CometSpec}, so for our CN fit, only lines in the range $\sim\!300\!-\!1000$~nm were included. It is important to highlight that solar UV variability affects the photodissociation \citep{Cochran93, Huebner15}, and hence, the production rates in comets. Therefore, the assumption of constant solar irradiance constitutes a systematic source of error, with no standard calibration procedure or solar monitoring to control for it.

For the spectral lines, we adopted the isotopolog line list provided by \cite{Brooke14} for \cni~and from \cite{Sneden14} for \cnii~and \cniii. Our analysis includes the \cni\ contributions of the $B^2\Sigma\!-\!X^2\Sigma$ $(\Delta v\!=\!0$; violet) system and the $A^2\Sigma - X^2\Sigma$ $(\Delta v\!=\!1$; red) system. For the latter, we only considered the $\Delta v\!=\!1$ transitions because the inclusion of the $A^2\Sigma - X^2\Sigma$ ($\Delta v \!=\!0, 2$) systems does not affect our fitted column densities, as we discuss in Sect.~\ref{sec:cnfit}. However, only the violet system was fitted in our analysis. This system alone lies within the ESPRESSO wavelength coverage.

These line lists provide all the information required to construct the fluorescence model. Nevertheless, \textsc{CometSpec} also allows the user to supply a custom line list. In this case, each line must include the wavelength, the Einstein spontaneous emission coefficient $A_{ul}$, an upper-state identifier, a lower-state identifier (which together define a unique transition, e.g., the notation $21$ for a transition from level 2 to level 1 in atomic systems), and the statistical weights of the upper and lower levels, $g_u$ and $g_l$. If rotational collisions are to be included, \textsc{CometSpec} also requires the lower electronic state, lower vibrational state, lower rotational quantum number $J$, lower-state symmetry (including the $F$ level and parity), and the energy of the lower state. This information ensures a unique identification of each transition and allows the computation of the energy difference $\Delta E_{ul}$ between rotational levels.

After all ingredients were specified, the fluorescence model was constructed by solving the system of Einstein rate equations. We defined the population vector
\begin{equation*}
x \equiv
\begin{pmatrix}
n_1 \
n_2 \
\dots \
n_N
\end{pmatrix}^T,
\end{equation*}

where $n_i$ is the fractional population of level $i$, and $N$ is the total number of energy levels included in the model. The total transition rate $P_{ij}$ from level $i$ to level $j$ is defined as
\begin{equation*}
P_{ij} \equiv R_{ij} + C_{ij},
\end{equation*}

where $R_{ij}$ represents radiative rates and $C_{ij}$ the collisional rates. The radiative rates are given by
\begin{equation*}
R_{lu}=B_{lu}J_\nu,
\end{equation*}
that is, absorption ($l \rightarrow u$), and
\begin{equation*}
R_{ul} = A_{ul} + B_{ul}J_\nu,
\end{equation*}
spontaneous and stimulated emission ($u \rightarrow l$). Here, $A_{ul}$ is the Einstein coefficient for spontaneous emission, and $B_{ul}$ and $B_{lu}$ are the Einstein coefficients for stimulated emission and absorption, respectively. $J_\nu(\nu_{ul})$ is the mean intensity of the radiation field evaluated at the transition frequency $\nu_{ul}$ (obtained from the solar irradiance seen by the comet), and $\nu_{ul}$ is the transition frequency, that is, the line frequency.

The collisional rates are related through
\begin{equation*}
g_l\,C_{lu} = g_u\,C_{ul}\,
\exp\left(-\frac{E_u - E_l}{kT_{\rm kin}}\right),
\end{equation*}
where $E_u$ and $E_l$ are the energies of the upper and lower levels involved in collisional coupling\footnote{Note that $E_u-E_l$ need not correspond to an observed emission line frequency, since collisional coupling also connects levels with forbidden or otherwise unobserved radiative transitions.}, respectively, $g_u$ and $g_l$ their statistical weights, $k$ is the Boltzmann constant, and $T_{\rm kin}$ is the kinetic temperature.  $T_{\rm kin}$ here is not a thermodynamic temperature as CN is not necessarily in thermodynamic equilibrium; we only assumed statistical equilibrium, including radiative and collisional processes, so $T_{\rm kin}$ should be taken as the velocity that characterizes the distribution of the colliding particles \citep{Manfroid09}.

The time-dependent equation for each level $i$, and assuming fluorescence equilibrium, is

\begin{equation*}
\frac{dn_i}{dt}
= \underbrace{\sum_{j \neq i} n_j P_{ij}}_{\text{into } i}
- \underbrace{n_i \sum_{j \neq i} P_{ji}}_{\text{out of } i}
= 0.
\end{equation*}
The indices $i$ and $j$ run over energy levels in the model, not just those connected by observed transitions. The system of equations couples combinations of levels found in the line list, with some pairs connected only by collisions, some only by radiative processes, and some by both. This system can be written in matrix form as
\begin{equation*}
M x = b,
\qquad
b \equiv
\begin{pmatrix}
1 \
0 \
\dots \
0
\end{pmatrix}^T.
\end{equation*}
The matrix $M$ is defined as
\begin{equation*}
M = 
\begin{pmatrix}
1 & 1 & \cdots & 1 \\[4pt]
R_{21}+C_{21} & -\sum\limits_{k \neq 2} (R_{k2}+C_{k2}) & \cdots & R_{2N}+C_{2N} \\[6pt]
\vdots & \vdots & \ddots & \vdots \\[4pt]
R_{N1}+C_{N1} & R_{N2}+C_{N2} & \cdots & -\sum\limits_{k \neq N} (R_{kN}+C_{kN})
\end{pmatrix}.
\end{equation*}
Along each row $i$, the off-diagonal elements $M_{ij}$ ($j \neq i$) are the rates into level $i$ from level $j$, while the diagonal element $M_{ii}$ is the (negative) total rate out of level $i$. A replacement of the first row of the matrix by unity and setting the first element of $b$ equal to 1 imposes the normalization condition
$\sum_i n_i = 1$, as in \cite{Magnani86} and \cite{Bromley21}. 

The inclusion of the collisional rates $C_{ul}$ introduces a large number of free parameters, following \cite{Manfroid09}. We therefore assumed $\log C_{ul} \equiv f_{\rm col}$ (with a single $f_{\rm col}$ value) and restricted collisional transitions to those with $\Delta J\!=\!0, \pm 1$. Together with the solar irradiance, the spontaneous emission coefficients $A_{ul}$ and the Einstein coefficient relations, these provide a complete set of ingredients to formulate the system of equations, which were solved numerically using a least-squares approach. We assumed an optically thin coma for the incoming solar irradiance and for the emitted fluorescence.

In summary, with (i) the solar irradiance spectrum at a given heliocentric velocity and heliocentric distance, (ii) a provided or default line list, and (iii) parameters such as fiber geometry, column density ($\log N$), line spread function (LSF), $f_{\rm col}$, and $T_{\rm kin}$ (if collisions are included), \textsc{CometSpec} allows the computation of a fluorescence emission model of one or multiple isotopologs (see App.~\ref{sec:CometSpec} for more details on the computation of uncertainties).

\subsection{CN fitting}\label{sec:cnfit}

\begin{figure}[t]
\centering
\includegraphics[width=\columnwidth]{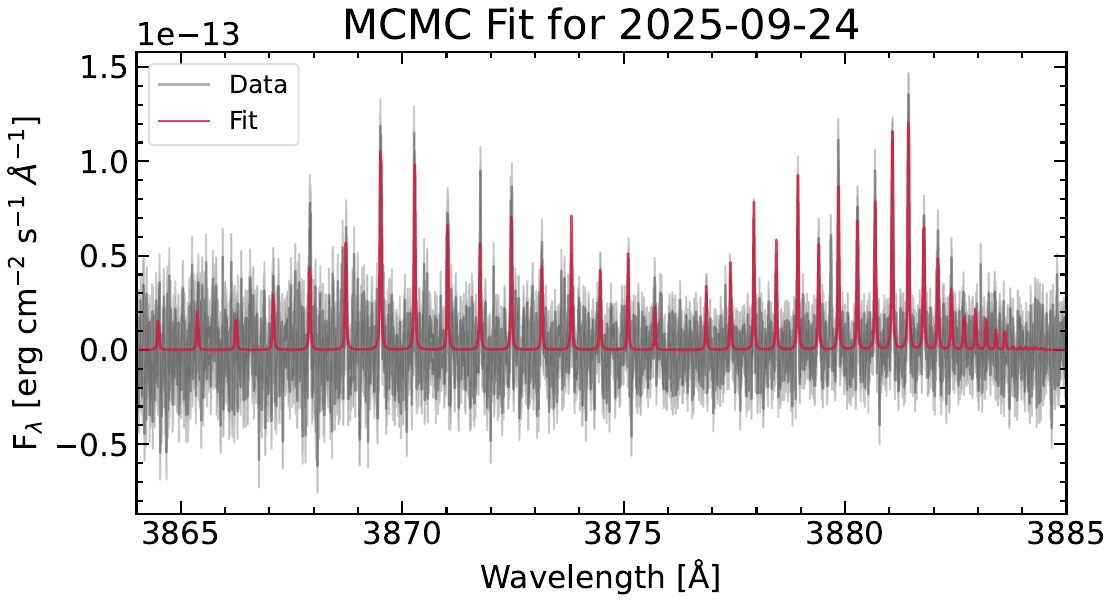}
\caption{Spectrum fit of the 3I violet CN band for the night of September 24. The gray line shows the observed data, and the shaded light gray region indicates the uncertainty. The red line represents the CN fluorescence model. \label{fig:cnfit}}
\end{figure}

We modeled the CN emission of the $B^2\Sigma - X^2\Sigma$ $(\Delta v\!=\!0$; violet) line system and fit the free parameters kinetic temperature $T_{\rm kin}$, the collision transition probability $ f_{\rm col}$, the column density $\log N$, the full width at half maximum ($\text{FWHM}_L$) of the Lorentzian profile LSF, and a velocity shift $\Delta v$ to account for residual mismatches between the observed lines and those from \cite{Brooke14}. The uniform priors on all parameters were $\log N\in[5, 20]$, $ f_{\rm col}\in[-8, 1]$ where the collisional rate is in units of s$^{-1}$, $T_{\rm kin}\in[10, 1\,000]$~K, $\text{FWHM}_L\in[0.001,0.03]$~\AA , and $\Delta v\in[-1, 1]$~km s$^{-1}$. All lines with $A_{ul}<10^4$ were filtered out, which substantially sped up the fitting and modeling. Throughout, \textsc{CometSpec} assumes input wavelength in \AA\ and flux densities in erg s$^{-1}$ cm$^{-2}$ \AA$^{-1}$.

The MCMC was run with a total of 30 walkers for $5\,000$ steps, sampling a parameter space of dimension five, producing $750\,000$ iterations in total. We applied the additional selection of the posterior samples (see App.~\ref{sec:CometSpec}). Fig.~\ref{fig:cnfit} shows an example of the fitting for the night of September 24, and the posterior distributions can be found in Fig.~\ref{fig:cnposterior}. Five aspects are worth pointing out from our fits. First, the line profile is not perfectly described by the model, so fitted lines can appear shallower due to a broader base. This is reflected as a correlation between $\log N$ and $\text{FWHM}_L$. Second, the relative line heights can deviate from the model, either because the Kurucz spectrum does not perfectly reproduce the solar irradiance at the time of observation or because of assumptions built into the fluorescence model (e.g., the $A_{ul}$ filtering). Third, the inclusion of collisions was essential for reproducing the observations with our adopted line set and filter configuration, as fits that neglected collisions showed significant discrepancies with the data. To assess whether this result was driven by the limited set of transitions included in the model, we performed an additional test of fitting three nights with good S/N by lowering the minimum $A_{ul}$ threshold to 10, extending the original set of systems included to $B^2\Sigma\!-\!X^2\Sigma$ $(\Delta v\!=\!0$; violet),  the $A^2\Sigma - X^2\Sigma$ $(\Delta v\!=\!0,1,2$; red), and the rovibrational transitions in the $X^2\Sigma^+$, and using the extended near-IR solar spectrum of \citet{Bromley24}. For this test, we decreased the iterations of the MCMC to $50\,000$. Under these conditions, the fitted values of $f_{\rm col}$ decreased (at the cost of high computational time), indicating that part of the collisional excitation inferred in the reference model compensated for the omission of transitions that contribute to the population of the violet system. Despite this reduction in $f_{\rm col}$, the retrieved column densities remained largely unchanged between the two model configurations. The analysis is therefore unaffected, although $f_{\rm col}$ itself should be interpreted with caution. Fourth, for some nights, the posterior distribution of $f_{\rm col}$ in the reference and IR-extended models exhibited a regime in which further increases in the collisional rate produced negligible changes in the predicted spectrum. This behavior suggests that beyond a certain threshold, collisions effectively thermalize the rotational population within the ground vibrational state. Fifth, deviations from the fits can also be explained when other excitation pathways in addition to fluorescence are considered. The fits and corner plots for all observing nights can be found online\footnote{\url{https://zenodo.org/records/20966879}}.

\subsection{Production rates}\label{sec:cnprod}

\textsc{CometSpec} can also compute production rates using standard Haser-model procedures \citep{Haser57}. The required inputs are the fluorescence model (specifically, the fitted column density), the gas outflow velocity, the parent and daughter scale lengths, and the aperture geometry. The spatial distribution is integrated within the aperture using \textsc{sbpy} \citep{Mommert19}, assuming the aperture is centered on the nucleus. We considered two sources of uncertainty on $\rm Q_{\rm CN}$: a statistical error from the MCMC fit, and a systematic error from slit losses.

The statistical error was obtained by propagating the column-density posterior chain into a production-rate chain and taking half the difference between the $84th$ and $16th$ percentiles. The slit-loss systematic accounts for flux that falls outside the fiber aperture under varying observing conditions. Given the minimum and maximum seeing and target altitude during a given night together with a reference wavelength and the aperture geometry, we estimated the corresponding range of sky point spread function (PSF) FWHM following \citet{Persson22}. Assuming a two-dimensional Gaussian PSF, we then computed the fraction of flux falling outside the aperture at each extreme, which maps directly into a range of inferred production rates via column densities ($\text{Q}_{\rm CN} \propto N \propto F$). The slit-loss error is defined as the mean of the absolute deviations of the nominal $\rm Q_{\rm CN}$ from these two extrema. The total uncertainty on $\rm Q_{CN}$ is the quadrature sum of the statistical and systematic contributions.

We adopted Solar System cometary scale lengths in the absence of 3I-specific constraints. Therefore, the derived $\rm Q_{\rm CN}$ values should be taken as relative rather than absolute, and a direct comparison with \cite{Rahatgaonkar25} is only valid to the extent that both works make the same assumption. We assumed an outflow velocity of $v\!=\!0.85\!\times\!r_{\rm h}^{-0.5}\;\text{km s}^{-1}$ \citep{Cochran93}, a parent scale length of $1.3\!\times\!10^4$\,km, and a daughter scale length of $2.1\!\times\!10^{5}$\,km scaled by $r^{-2}$ at 1\,au \citep{AHearn95}. We stress, however, that 3I has an unusually CO$_2$- and CO-rich coma where the outflow dynamics might differ. For instance, \citet{Cordiner26} measured an HCN outflow velocity of $v = 0.276 \pm 0.015$\,km s$^{-1}$, which is lower than the $0.551$\,km s$^{-1}$ expected from the \citet{Cochran93} relation at the corresponding heliocentric distance. Adopting the latter would yield a production rate twice as high as the one derived from the \citet{Cordiner26} velocity. Similarly, the HCN $\to$ CN scale lengths from \cite{AHearn95} assume solar-driven HCN photodissociation rates, which is likely appropriate for 3I, but should be kept in mind as a potential systematic.

\begin{figure}[h!]
\centering
\includegraphics[width=1\columnwidth]{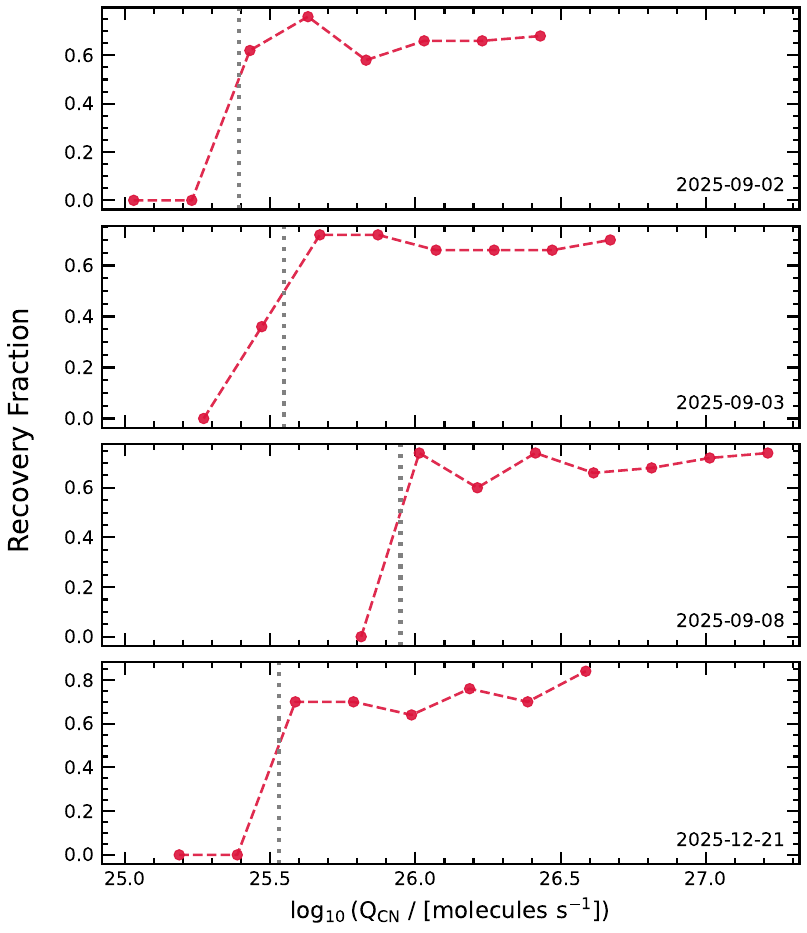}
\caption{Recovery fraction as a function of CN production rate for the nights of September 2, 3, and 8, and December 21. The red points show the values obtained as described in Sect.~\ref{sec:cnprod}, and the dashed red line shows a linear interpolation. The dotted black line indicates the adopted upper limit, as described in Sect.~\ref{sec:cnprod}.\label{fig:cnupperlim}}
\end{figure}

For the nights of September~2, 3, 8, and December~21, CN was not clearly detected, and we instead derived 50\% completeness upper limits as follows (Fig.~\ref{fig:cnupperlim}). For each night, we generated 50 synthetic spectra matching the observed signal-to-noise ratio conditions, spanning a range of injected CN column densities. Each realization was fitted with an MCMC procedure analogous to that described above, with three simplifications: the iterative posterior filter was disabled, we used $6\,000$ iterations per fit, and only $\log N$ was left free, with a broad uniform prior. All the other parameters were set to a first fit of the few lines detected each night that followed the same procedure explained in Sect.~\ref{sec:cnfit}. A recovery was counted as successful when (i) the injected column density fell within the $1\sigma$ uncertainty interval of the recovered value and (ii) the relative uncertainty on $N$ was below 10\%. The recovery fraction at each injected column density was then the number of successful recoveries divided by 50. We converted the injected column-density grid into production rates and defined the upper limit as the interpolated $\rm Q_{\rm CN}$ value at which the recovery fraction first dropped to 50\%; the slit-loss error was included. Table~\ref{tab:results} reports the resulting $\log \rm Q_{\rm CN}$ for each nondetection and detection night.

\begin{figure*}[t!]
\centering
\includegraphics[width=2\columnwidth]{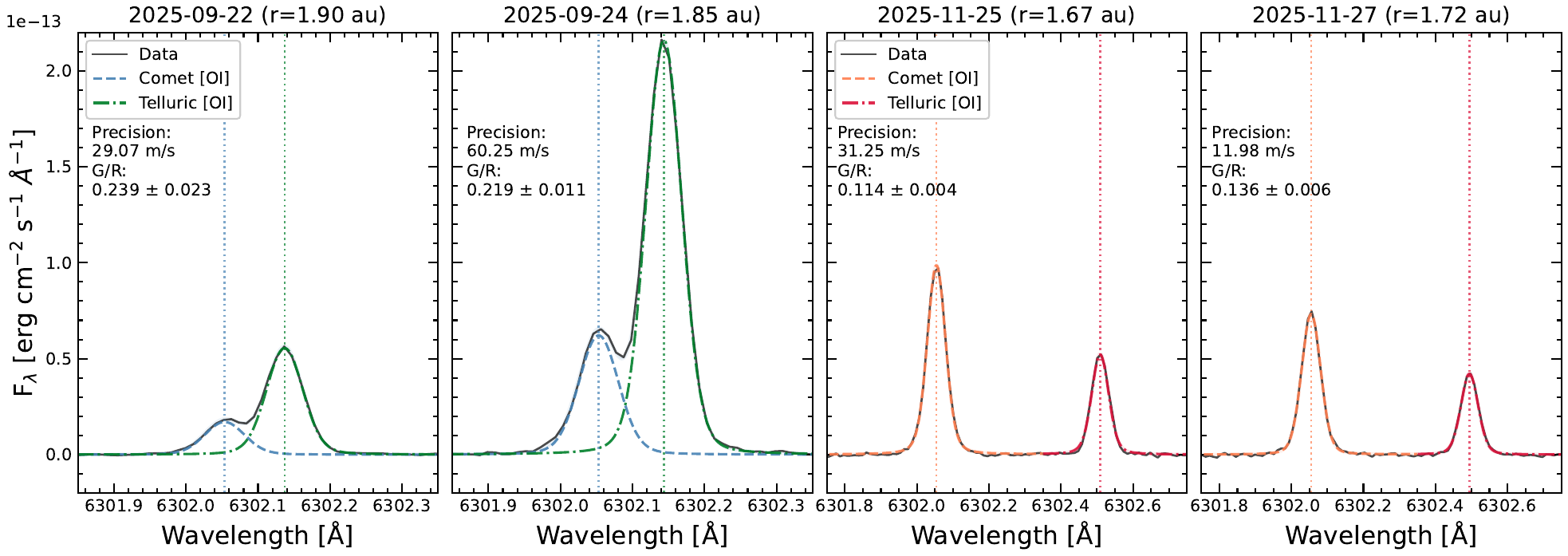}
\caption{Fits of the [\oi] 6302\,\AA\ line for four nights: September 22 ($r_{\rm h}\!=\!1.9$\,au) and 24 ($r_{\rm h}\!=\!1.85$\,au) pre-perihelion (left panels) and November 25 ($r_{\rm h}\!=\!1.67$\,au) and November 27 ($r_{\rm h}\!=\!1.72$\,au) post-perihelion (right panels). The solid black line shows the observed data. The cometary line fit is shown in dashed blue (pre-perihelion) and orange (post-perihelion), while the telluric fit is shown in dot-dashed green (pre-perihelion) and red (post-perihelion). The G/R ratio is indicated in each panel, along with the precision estimated as the difference of $\dot{\Delta}$ estimated from the fit and the ephemeris value. \label{fig:oxfit}}
\end{figure*}

\subsection{Forbidden oxygen}\label{sec:oxygen}

Each forbidden oxygen line at 5578, 6302, and 6365~\AA\ was modeled as the sum of two Voigt profiles, one for the telluric component, and the other for the cometary emission, with the Lorentzian FWHM tied between the two. Instead of fitting the lines independently, we fitted them jointly in a single MCMC run over a 21-dimensional parameter space, comprising three shared $\mathrm{FWHM}_L$ (one per line) and six values each of $\mathrm{FWHM}_G$, the Lorentzian amplitude $A_L$, and the line center $x_0$ (one per Voigt component). The sampler was run with 100 walkers for $5\,000$ steps, totaling $1.05\times10^{7}$ iterations, and the iterative posterior filter described in App.~\ref{sec:CometSpec} was again applied.

Priors were tuned per night to ensure convergence and were highly restrictive in some cases. This did not bias our analysis because the quantities of astrophysical interest are the line intensities and centers. We prioritized obtaining accurate line profiles, in which the telluric and cometary components are clearly separated, and accepted larger uncertainties on the remaining (nuisance) parameters in the few highly blended cases where degeneracies dominate.

Figure~\ref{fig:oxfit} shows four representative fits of the [O\,\textsc{i}]~6302\,\AA\ line, two pre-perihelion and two post-perihelion. The full set of fits and corner plots is available online\footnote{\url{https://zenodo.org/records/20966879}}. Table~\ref{tab:results} summarizes the main results of the [O\,\textsc{i}] fits.

\subsection{Geocentric radial velocity}\label{sec:geocentric}

\begin{figure}[h]
\centering
\includegraphics[width=\columnwidth]{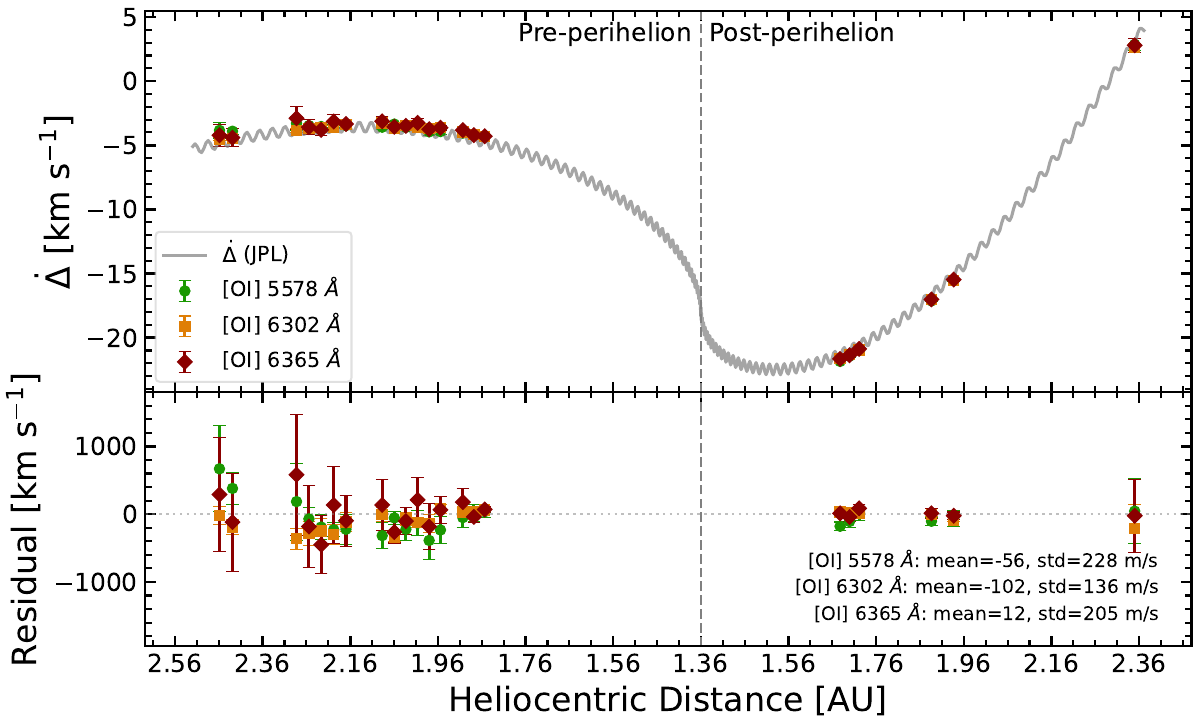}
\caption{Geocentric 3I velocity ($\dot{\Delta}$) derived from the forbidden oxygen lines by differencing the cometary and telluric [O\,\textsc{i}] line centroids. {\it Upper panel:} Per-line measurements, with green circles for [O\,\textsc{i}]~5578\,\AA, orange squares for [O\,\textsc{i}]~6302\,\AA, and red diamonds for [O\,\textsc{i}]~6365\,\AA. The gray line shows the JPL Horizons ephemeris value. {\it Lower panel:} Velocity residuals relative to the ephemeris.     \label{fig:dv}} 
\end{figure}

The high spectral resolution of ESPRESSO, combined with the data quality, allowed us to clearly separate the telluric and cometary lines on nearly every night. As anticipated in Sect.~\ref{sec:red}, anchoring the wavelength solution to the [O\,\textsc{i}]~5578\,\AA\ telluric line and applying a $\dot{\Delta}$ correction yielded a sufficient precision for stacking within and across nights. This is confirmed by the close agreement between the ephemeris velocity and the velocity derived from the cometary-telluric centroid offset. The residuals fell below $1\,\mathrm{km\,s^{-1}}$ on all nights and reached $\sigma\simeq\!135\!-\!228\,\mathrm{m\,s^{-1}}$. Figure~\ref{fig:dv} illustrates this remarkable agreement.

\begin{figure*}[t]
\centering
\includegraphics[width=2\columnwidth]{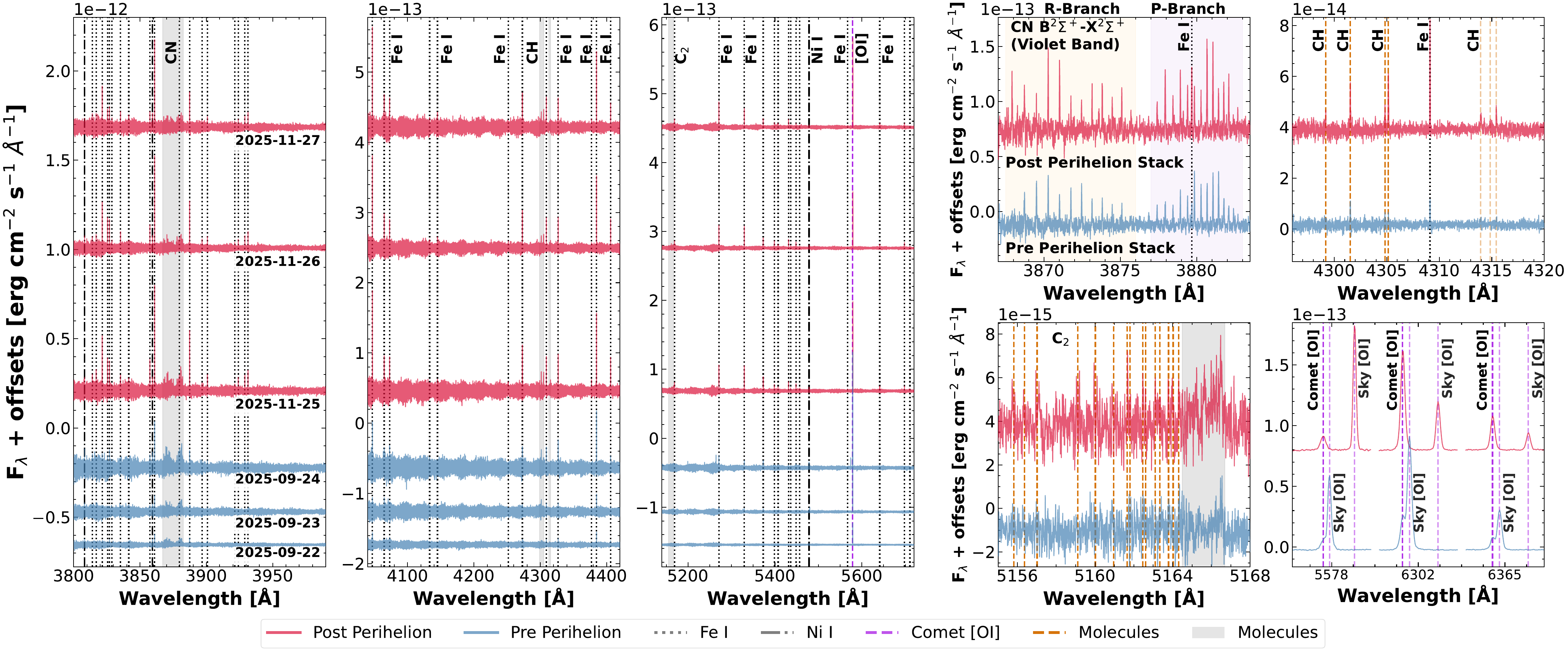}
\caption{Detected emission features in the ESPRESSO spectra of 3I/ATLAS for pre- and post-perihelion stacks (blue and red spectra, respectively) using individual spectra from six nights (three pre-perihelion: September 22, September 23, and September 24 in blue; three post-perihelion: November 25, November 26, and November 27 in red). \textit{Three left columns:} Coverage of the ranges 3800--3950\,\AA, 4100--4400\,\AA, and 5200--5700\,\AA, with spectra offset vertically for clarity. {\it Two right columns:} Pre- and post-perihelion stacks of the same nights, showing the CN $B^2\Sigma^+$--$X^2\Sigma^+$ violet band and its $R$ and $P$ branches (3867--3882\,\AA), the CH band (4300--4320\,\AA), the C$_2$ Swan band (5156--5166\,\AA), and the three cometary forbidden [O\,\textsc{i}] lines at 5578, 6302, and 6365\,\AA\ alongside their telluric counterparts. In all panels, the vertical dashed lines and shaded bands mark the positions of the identified species that are labeled in the legend.\label{fig:prepost}} 
\end{figure*}

\subsection{Detected emission lines}\label{sec:lines}

\begin{figure*}[t]
\centering
\includegraphics[width=2\columnwidth]{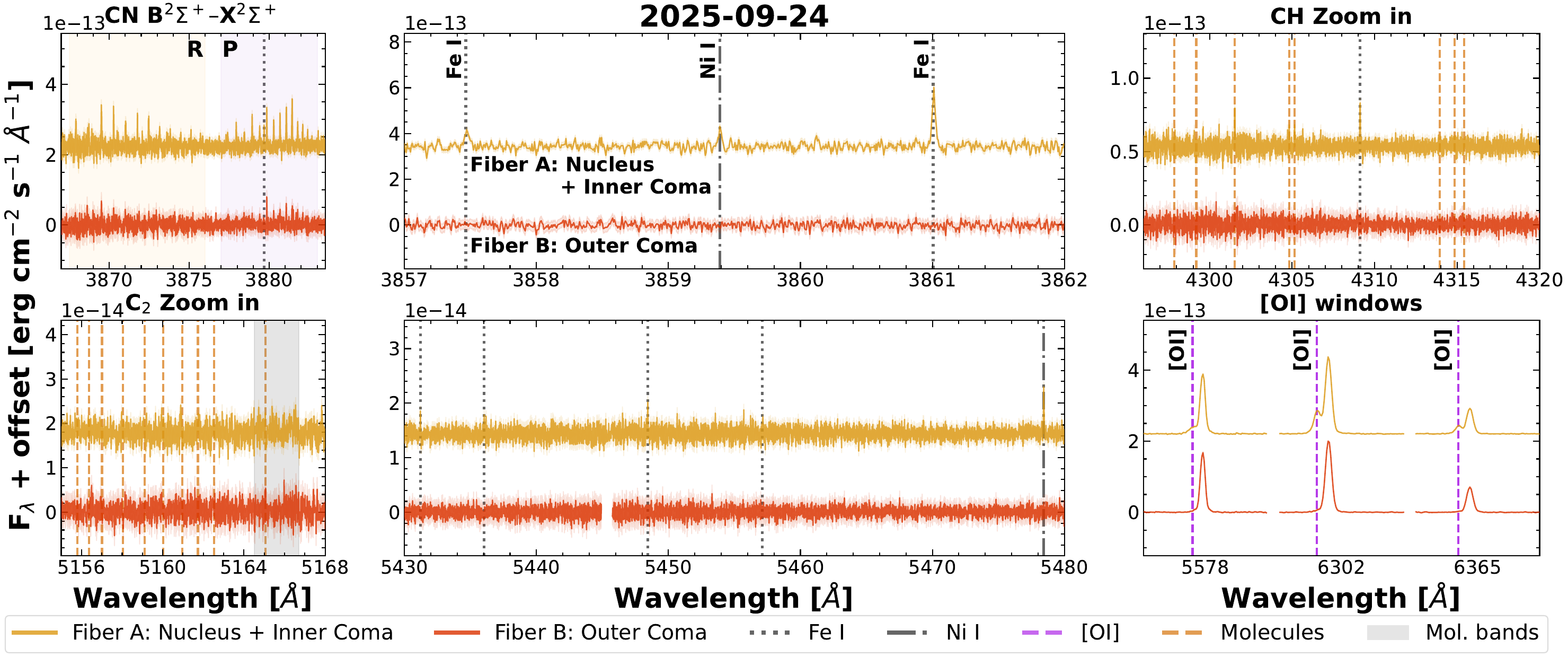}
\caption{Comparison of fiber A (nucleus; yellow) and fiber B (coma; orange) spectra obtained on  September 24, covering key spectral regions: the CN $B^2\Sigma^+$--$X^2\Sigma^+$ violet band with its R and P branches (3867--3882~\AA), Fe\,\textsc{i} and Ni\,\textsc{i} lines (3857--3862~\AA\ and 5430--5480~\AA), the CH band (4300--4320~\AA), the C$_2$ Swan band ($5156$--5166~\AA), and the three [\oi] forbidden lines at 5578, 6302, and 6365~\AA. In all panels, the vertical dotted, dashed, and dot-dashed lines indicate the positions of Fe\,\textsc{i}, Ni\,\textsc{i}, [\oi], and molecular lines, respectively, with shaded bands marking molecular band regions as labeled in the legend. In this night, fiber B was placed at a radial distance from the nucleus of $\sim\!12\,800$~km. \label{fig:ab}} 
\end{figure*}

In addition to the violet CN band and the forbidden oxygen lines, the ESPRESSO spectra reveal several additional emission features (see Fig.~\ref{fig:prepost}). Atomic iron and nickel lines were identified from the National Institute of Standards and Technology (NIST) Atomic Spectra Database \citep{NIST_ASD}, and molecular features were identified using the cometary emission line catalog of \citet{Cambianica21}. To characterize the evolution of these features across the perihelion passage of 3I and to probe their spatial distribution within the coma, we carried out two comparisons: pre- versus post-perihelion stacked spectra (Fig.~\ref{fig:prepost}), and fiber~A (nucleus-centered) versus fiber~B (projected offset) detections in individual nights (Fig.~\ref{fig:ab}). We also performed a line-stacking search for isotopologs, which yielded no detection (see App.~\ref{sec:isotop} for details).

\subsubsection{Pre- versus post-perihelion evolution} 
The pre and post comparison reveals two notable changes. First, we found a post-perihelion enhancement of Fe\,\textsc{i} emission when comparing the spectra in Fig.~\ref{fig:prepost}. This difference is mainly explained by the slightly smaller heliocentric distances ($\sim\!0.17$~au) of the post-perihelion observations, which enhance the Fe lines. A direct comparison between nights E17 ($1.85$~au) and E21 ($1.88$~au) yields production rates $\log\text{Q}_{\rm Fe}=24.42 \pm 0.06$ and $\log\text{Q}_{\rm Fe}=24.41 \pm 0.08$, respectively ($\rm Q$ in molecules s$^{-1}$), showing no clear asymmetry, although a robust assessment of the Fe pre- and post-perihelion asymmetric behavior predicted by thermal models requires comparing a larger sample of nights, which lies beyond the scope of this work (see Sect.~\ref{sec:cnprod} for details on the production rate derivation and App.~\ref{app:fec2ch} for the Fe case). When comparing these two values to those presented by \cite{Hutsemekers26}, we found that for similar heliocentric distances, our values are higher by $\sim\!0.3$ dex. These differences can be explained by a different set of lines used for fluorescence calculations (see App.~\ref{app:fec2ch} for the specific differences between the models), imperfections in the LSF treatment due to different prescriptions of simultaneous lines fits, centering offsets during observations, or aperture effects because our fiber is smaller than the Ultraviolet-Visible Echelle Spectrograph (UVES) and X-Shooter slits. Second, C$_2$ (Swan bands) and CH (4300\,\AA\ band) are detected more clearly in the post-perihelion stack than in the pre-perihelion data. From the stack of the first three post-perihelion nights, we derived $\log \rm Q_{C_2}$ in the range of $[25.20,25.68]$, corresponding to $\log \rm Q_{C_2}/Q_{CN}=[-0.79,-0.10]$ when compared to the mean CN production rate over the same epochs (see App.~\ref{app:fec2ch} and Sect.~\ref{sec:cnprod} for details on the line lists used and the fitting procedures for Fe, $\rm C_2$, and CH). This ratio places 3I within or near the boundary of the carbon-chain-depleted category defined by \cite{AHearn95}, where $\log \rm Q_{C_2}/Q_{CN}<-0.18$. Our measurement is broadly consistent with the pre-perihelion classifications of \citet{Rahatgaonkar25}, \citet{Salazar25}, \citet{Lazzarin26}, and \citet{Hutsemekers26}, although 3I does not appear to represent an extreme case of carbon-chain depletion. Our $\log \rm Q_{C_2}$ values agree with those reported by \citet{Kawakita26}; nevertheless, our range remains consistent with carbon-chain depleted classification. We caution, however, that our $\rm C_2$ model is subject to high degeneracies, and a systematic underestimation of the associated uncertainties cannot be excluded (see App.~\ref{app:fec2ch}). In addition, the limit of $-0.18$ defined by \citet{AHearn95} to classify a comet as $\rm C_2$-depleted comes from narrow-band photometry, which implies fields of view (FoVs) larger than our fiber. The aperture effects are therefore another point requiring caution. For CH, the same post-perihelion stack yielded $\log \rm Q_{CH} = 26.38 \pm 0.05$ and $\log \rm Q_{CH}/Q_{CN}=0.31\pm0.12$, which are both within the range of typical Solar System comets \citep{Cochran12}.

\subsubsection{Spatial systematics from the radially offset fiber B}\label{sec:spatial}
Not every epoch is suitable for the nucleus versus coma (i.e., fiber~A versus fiber~B) comparison because the position angle of fiber~B rotates on the sky between exposures, and depending on which UT is used, more or less coma flux enters fiber~B (see Sect.~\ref{sec:obs}). On September 24, 2025, however, the two fibers yielded significant detections: CN is clearly present in fiber~B at a projected nucleocentric distance of $\sim\!12\,800$\,km, while Fe\,\textsc{i} and Ni\,\textsc{i} emissions were absent at such large nucleocentric distances, consistent with their known origin close to the nucleus \citep{Hutsemekers21, Manfroid21, Rahatgaonkar25, Hutsemekers26}.

As an independent check on the adopted CN parent scale length, we compared the column density predicted by the Haser model at the projected nucleocentric distance of fiber~B ($7\arcsec$) with the column density obtained from a fluorescence fit to the fiber~B spectrum (see Sect.~\ref{sec:cnfit} for fitting details). The Haser prediction yields $\log N/(\text{molecules\,cm}^{-2}) = 11.31 \pm 0.06$, while the fiber~B fit gives $\log N/(\text{molecules\,cm}^{-2}) = 11.23 \pm 0.07$, which agrees very well. This consistency supports the validity of the adopted HCN--CN scale lengths and outflow velocity to compute the production rates (see Sect.~\ref{sec:cnprod}).

\begin{figure*}[htp!]
\centering
\includegraphics[width=1.8\columnwidth]{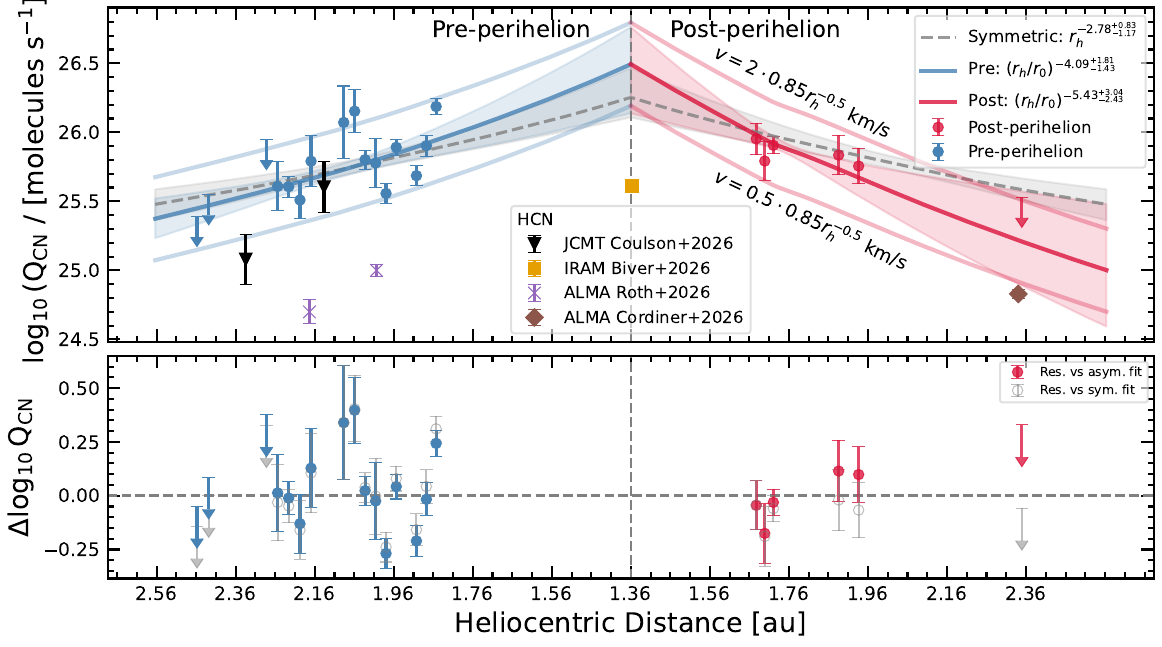}
\caption{CN evolution during the ESPRESSO campaign. {\it Upper panel:} CN production rate $\rm Q_{\rm CN}$ (logarithmic scale) as a function of heliocentric distance $r_{\rm h}$. The blue and red symbols denote pre- and post-perihelion measurements, with $1\sigma$ error bars and downward arrows for upper limits. The vertical dashed line marks perihelion. The dashed black line and gray band show the symmetric fit $\text{Q}_{\rm CN}\!=\!a\,r_{\rm h}^{\,b}$ and its $1\sigma$ confidence region. The blue and red curves show the asymmetric piecewise fit, with continuity at perihelion. No upper limits are in the fit. The two lighter solid lines show the effect of doubling and halving the adopted outflow velocity on the fitted power law, bracketing the systematic uncertainty introduced by the model assumptions. The black triangles, the yellow square, the violet crosses, and the brown diamond show HCN production rates from \citet{Coulson26, Biver26, Roth26b, Cordiner26}, respectively. {\it Lower panel:} Residuals with respect to the asymmetric fit (filled symbols, color-coded as above) and to the symmetric fit (open gray symbols). The horizontal dashed line marks the zero residual.\label{fig:cnevol}}
\end{figure*}

\section{Discussion}
\label{sec:disc}

\subsection{Evolution of the CN production rate}

\begin{table*}[h]
\centering
\caption{CN fits of the heliocentric scaling relation $\text{Q}_{\rm CN}=a\,(r_{\rm h}/1\text{ au})^{\,b}$ and $\text{Q}_{\rm CN} = \text{Q}_{\rm peri}\,(r_{\rm h}/r_0)^{\,b}$}
\label{tab:cn_fits}
\setlength{\extrarowheight}{2pt}
\begin{tabular}{ccccccc}
\hline\hline
Data included & \multicolumn{3}{c}{Upper limits included} & \multicolumn{3}{c}{Upper limits not included} \\
 & $\log a$ & $b$ & $\log \mathrm{Q_{peri}}$ & $\log a$ & $b$ & $\log \mathrm{Q_{peri}}$ \\
\hline
Both & $26.84^{+0.33}_{-0.22}$ & $-3.59^{+0.70}_{-1.01}$ & --- & $26.62^{+0.37}_{-0.25}$ & $-2.78^{+0.83}_{-1.17}$ & --- \\
Pre  & --- & $-4.62^{+1.25}_{-1.22}$ & $26.57^{+0.24}_{-0.26}$ & --- & $-4.09^{+1.81}_{-1.43}$ & $26.49^{+0.27}_{-0.35}$ \\
Post & --- & $-6.14^{+2.16}_{-2.18}$ & $26.57^{+0.24}_{-0.26}$ & --- & $-5.43^{+3.04}_{-2.43}$ & $26.49^{+0.27}_{-0.35}$\vspace{0.5mm}\\
\hline\hline
\end{tabular}
\end{table*}

\begin{table*}[h!]
\centering
\caption{Power-law fits of the heliocentric scaling relation $F = c\,(r_{\rm h}/r_0)^{a}$ for the [\oi] forbidden lines.}
\label{tab:OI_fits}
\begin{tabular}{lccc}
\hline\hline
Parameter & [\oi]~5578~\AA & [\oi]~6302~\AA & [\oi]~6365~\AA \\
\hline
$a$ (Pre) & $-4.52^{+0.74}_{-0.79}$  & $-7.22^{+0.84}_{-1.01}$  & $-7.33^{+0.74}_{-1.11}$ \\
$a$ (Post)& $-3.35^{+1.35}_{-1.71}$  & $-5.51^{+1.43}_{-1.77}$  & $-5.67^{+0.99}_{-1.62}$ \\
$c$ (erg\,cm$^{-2}$\,s$^{-1}$)& $(2.24^{+0.96}_{-0.60})\times10^{-15}$ & $(2.05^{+1.11}_{-0.58})\times10^{-14}$ & $(7.40^{+3.35}_{-1.62})\times10^{-15}$ \vspace{0.5mm} \\
\hline\hline
\end{tabular}
\end{table*}

We modeled the heliocentric evolution of $\rm Q_{\rm CN}$ with two power laws: a symmetric form $\text{Q}_{\rm CN}=a\,(r_{\rm h}/1\text{ au})^{\,b}$ fitted jointly to the pre- and post-perihelion data, and an asymmetric piecewise form $\text{Q}_{\rm CN} = \text{Q}_{\rm peri}\,(r_{\rm h}/r_0)^{\,b}$ with independent exponents on each branch and continuity enforced at perihelion ($r_0 = 1.36$~au). The two fits were performed in $\log \rm Q_{\rm CN}$ space, minimizing a negative Gaussian log-likelihood with the Nelder-Mead algorithm, performed with and without upper limits as censored data. The best-fit values and $1\sigma$ uncertainties correspond to the $16th$, $50th$, and $84th$ percentiles of a $10\,000$-iteration bootstrap Monte Carlo. Figure~\ref{fig:cnevol} shows the resulting $\text{Q}_{\rm CN}(r_{\rm h})$ scaling relations (see Table~\ref{tab:cn_fits} for fit results) and the fit residuals. The pre-perihelion power-law exponent derived here is shallower than the one reported by \citet{Rahatgaonkar25}. We attribute the difference to four factors: (i) the substantial epoch-to-epoch scatter in $\rm Q_{\rm CN}$, which broadens the confidence region of the fit, (ii) the absence of detections at our largest sampled heliocentric distances, where only upper limits are available, (iii) the inclusion of slit-loss uncertainties, which were not accounted for by \citet{Rahatgaonkar25}, and (iv) aperture effects. All exponents we derived remain within the range observed for Solar System comets \citep{AHearn95}.

We assessed the relative merit of the symmetric and asymmetric CN power-law models using the Bayesian information criterion (BIC). When only detections were included, the asymmetric model was not preferred ($\Delta\mathrm{BIC}\!=\!3.5$), primarily because the sparse sampling at large heliocentric distances leaves the post-perihelion branch poorly constrained. When upper limits were included, however, the asymmetric model was mildly favored ($\Delta\mathrm{BIC}\!=\!6.9$), consistent with thermal models that predict pre- and post-perihelion asymmetries arising from the delayed heating of subsurface layers \citep{Puzia25, Yaginuma26}, a behavior also observed in Solar System comets \citep{Schleicher91}. However, the pre- and post-perihelion exponents are consistent within the errors, which implies that more sampling is needed to robustly confirm the asymmetry. Unexpectedly, the best-fit post-perihelion is steeper than the pre-perihelion slope. Thermal-lag arguments predict the opposite: residual subsurface heating after perihelion should sustain HCN sublimation and produce a shallower decline in Q$_{\rm CN}$ on the outbound branch. Because the post-perihelion data are consistent with a single power-law decline and the limited $r_{\rm h}$ coverage after perihelion prevents a robust independent measurement of the post-perihelion exponent, we regard this apparent steepening as driven by the sparse sampling rather than by a physical effect. The high bootstrap error of the fitted exponents underscores the importance of dense temporal sampling and sustained monitoring across the full perihelion passage of such objects. We note, however, that all variants of our fit (pre- and post-perihelion only or both branches, with or without upper limits) yield exponents that are mutually consistent within their statistical and systematic uncertainties. Two refinements that are beyond the scope of the present work might further improve the CN evolution model: (i) allowing the peak production rate to be offset from perihelion, as was observed in some Solar System comets \citep{Knight13}, and (ii) incorporating an explicit temperature dependence through an Arrhenius sublimation term.

The lower panel of Fig.~\ref{fig:cnevol} shows substantial scatter in the residuals, indicating that the simple power-law models do not capture the full variability of $\rm Q_{\rm CN}$. We considered several candidate drivers of this dispersion. Daily variations in the far-ultraviolet (FUV) irradiance cannot be neglected, as they directly affect the HCN photodissociation rate \citep{Huebner15}. We integrated the daily irradiance from the flare irradiance spectral model 2 (FISM2\footnote{\url{https://lasp.colorado.edu/lisird/data/fism_daily_hr}}) over the $0.01$--$100$\,nm range and found that during our observations, the standard deviation was $\sim\!10$\,\% of the mean total FUV flux, suggesting that solar UV variability might contribute to the observed scatter. Nucleus rotation \citep{Schleicher90} can drive single-period modulation; phase-folding the residuals at the currently published rotation periods of 3I, $P_{\rm rot}\!=\!15.48$\,h \citep{Serra26}, $16.16$\,h \citep{Santana25}, or $16.79$\,h \citep{Fuente25}, revealed no coherent pattern, and a Lomb--Scargle periodogram of the $\rm Q_{\rm CN}$ residuals showed no significant periodicity within a 24-hour search window. We cannot rule out that rotational modulation contributes to the observed scatter at a level below our detection threshold, however. Localized spontaneous outbursts might contribute to the scatter, but resolved photometry is needed. The solar wind is another plausible contributor because it might alter the chemical environment within the coma. The cometary response to coronal mass ejection events was studied in the UV \citep{Noonan18, Gotz24}, but analogous investigations in the optical relating emission to solar wind velocity, density, and temperature on daily timescales are still lacking. Target-of-opportunity campaigns triggered by heliospheric modeling would help fill this gap.

More significantly, systematic effects from the assumed Haser model might lead to an under- or overestimation of the production rate. In Fig.~\ref{fig:cnevol} we show two lines corresponding to outflow velocities scaled by factors of two and one-half (the latter matching the outflow velocity measured by \citet{Cordiner26} when $0.85\,r_{\rm h}^{-0.5}$ is rescaled). The relative values remain consistent, suggesting systematics as a plausible cause for differences between datasets and studies. For instance, if the outflow velocity varies with heliocentric distance in a nontrivial manner deviating from $r^{-0.5}$ scaling relation, the elevated production rates might be explained solely by an outflow velocity that is higher than assumed. It was also shown that an under- or overestimation of daughter and parent scale lengths can lead to significant production rate differences \citep{Fray05}. Another potential systematic is the seeing conditions during the night. In our case, the maximum seeing values in each night correlate mildly with the $\rm Q_{\rm CN}$ residuals, albeit without clear statistical significance (Pearson $r = 0.47$, $p = 0.05$). We therefore do not consider this a major effect because it is already incorporated into our analysis via the slit-loss error. Finally, comet miscentering during exposures represents a potential source of scatter. Re-centering was required between each exposure, and positional deviations within a single exposure, combined with the direction of drift, might contribute to this effect. To assess this systematic, we examined the continuum evolution and the residuals (see App.~\ref{sec:apcont}), which again exhibited significant scatter. All the possible explanations for the high scatter proposed so far need to be analyzed by taking into account that the fiber size might also amplify these effects.

\begin{figure*}[ht!]
\centering
\includegraphics[width=1.8\columnwidth]{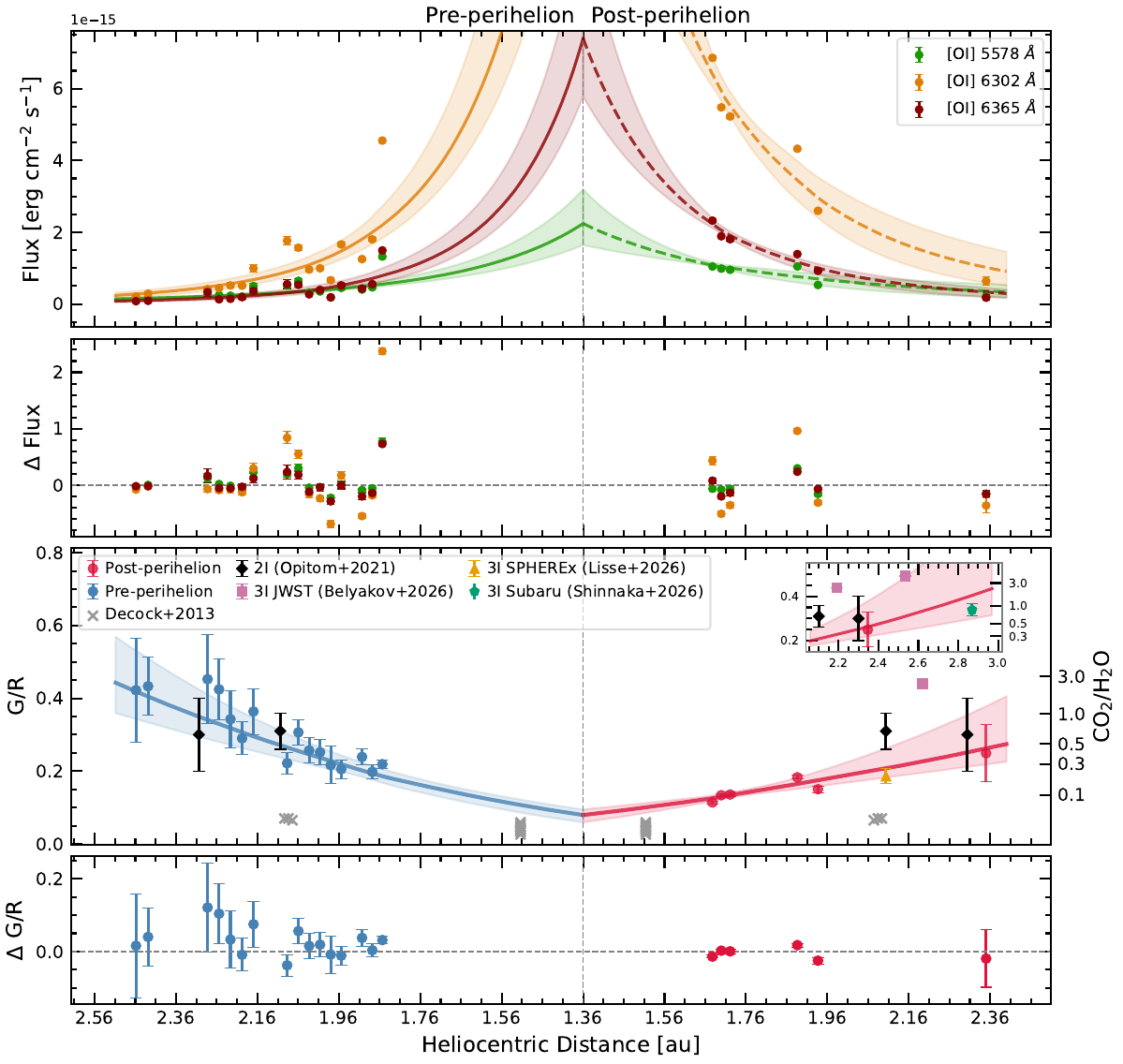}
\caption{3I oxygen forbidden line fluxes and the green-to-red ratio (G/R) as a function of heliocentric distance. \textit{Top panel:} Individual \text{[\oi]} line fluxes (5578~\AA\ in green, 6302~\AA\ in orange, and 6365~\AA\ in dark red) with asymmetric power-law fits of the form $c(r_{\rm h}/r_0)^a$, allowing independent slopes for the pre- and post-perihelion branches while enforcing continuity at perihelion ($r_0 = 1.36$~au). The solid lines show the pre-perihelion model, and the dashed lines show the post-perihelion model, each in the corresponding line color. The shaded regions indicate the 16th--84th percentile confidence bands. \textit{Second panel:} Residuals of each line flux relative to its corresponding fit, color-coded as above. \textit{Third panel:} G/R ratio, defined as the \text{[\oi]}~5578~\AA\ flux divided by the sum of the \text{[\oi]}~6302~\AA\ and 6365~\AA\ fluxes. Our data are shown as colored circles (teal for pre-perihelion, and crimson for post-perihelion). The solid curves and confidence bands of the G/R model were computed from the ratio of the individual line fits, with uncertainties propagated through bootstrap resampling. The right y-axis shows the estimate of $\text{CO}_2/\text{H}_2\text{O}$ using the \cite{Decock13} prescription. The inset plot in the top right corner corresponds to an extension of the G/R fitted model to larger heliocentric distances, with the same axes. The gray crosses correspond to comets 9P/Tempel 1 (Jupiter family) and C/2009 P1 Garradd ($a\!>\!10\,000$~au) from \cite{Decock13} and black crosses to 2I/Borisov from \cite{Opitom21}, both of which are duplicated in the pre- and post-perihelion panels. The pink squares show measurements from \citet{Belyakov26}, and the yellow triangle shows a measurement from \citet{Lisse26}, both corresponding to post-perihelion \coii/\hiio\ measurements. The green pentagon shows the G/R ratio measured by \citet{Shinnaka26}. \textit{Bottom panel:} Residuals of the G/R data relative to the derived model, with the same color-coding. In all panels, the dashed vertical gray line marks perihelion, and the horizontal dashed line in the residual panels shows the zero-residual value.\label{fig:oxevol}}
\end{figure*}

Since CN in cometary comae is primarily produced through the photodissociation of HCN, simultaneous optical (CN) and submillimeter (HCN) observations are essential to distinguish whether the observed scatter reflects true physical variability in the coma environment or residual systematics. Such coordinated campaigns would additionally clarify whether HCN is the sole parent driving CN production in 3I or if additional sources (e.g., CN-bearing grains or other parents) contribute, as was observed for some Solar System comets \citep{Fray05}. These considerations should be kept in mind for future interstellar visitor (and comet) observing campaigns, where coordinated multiwavelength campaigns from discovery onward will be essential to maximize the scientific return from these rare objects. From the literature, we retrieved 3I measurements of HCN from three facilities: James Clerk Maxwell Telescope \citep[JCMT,][]{Coulson26}, Instituto de Radioastronomía Milimétrica \citep[IRAM,][]{Biver26}, and the Atacama Large Millimeter/submillimeter Array \citep[ALMA,][]{Roth26b, Cordiner26}. They are all shown in Fig.~\ref{fig:cnevol}. Comparing these values with our asymmetric fit (without upper limits), and scaling the outflow velocity to that of \citet{Cordiner26}, we found $\log(\rm Q_{CN}/Q_{HCN}) = 0.15\pm0.20$ and $-0.22\pm0.19$ for \citet{Coulson26}, $0.66\pm0.10$ and $0.50\pm0.07$ for \citet{Roth26b}, $0.57\pm0.35$ for \citet{Biver26}, and $0.09\pm0.36$ for \citet{Cordiner26}. The values from \citet{Coulson26} and \citet{Cordiner26} are consistent with HCN being the dominant parent of CN, whereas those from \citet{Roth26b} and \citet{Biver26} are not. With the current data and differing analysis methods, we cannot determine which is correct, mainly because the effects of aperture, beam size, heliocentric distance, and sparse sampling are hard to separate.

\subsection{G/R ratio}
We applied the same fitting procedure used for $\rm Q_{\rm CN}$ to each forbidden [O\,\textsc{i}] line flux, modeling its heliocentric distance dependence as a piecewise power law, $F\!=\!c \left(r_{\rm h}/r_0\right)^a$, with independent pre- and post-perihelion exponents and continuity enforced at perihelion. Table~\ref{tab:OI_fits} summarizes the best-fit parameters and their $1\sigma$ uncertainties from the bootstrap experiments.

The results are presented in Fig.~\ref{fig:oxevol}. The upper panels show the flux of the three forbidden [O\,\textsc{i}] lines as a function of heliocentric distance, together with the corresponding piecewise power-law fits and residuals. The lower panels show the green-to-red ratio (G/R) and its residuals. The G/R model and its confidence band were derived from the ratio of the individual line fits, evaluated at each bootstrap iteration to propagate the correlated uncertainties.

The green [O\,\textsc{i}]~5578\,\AA\ line is produced primarily by the photodissociation of \hiio\ and \coii, while the red doublet at 6302 and 6365\,\AA\ originates predominantly from H$_2$O \citep{Fink84, Combi91}, with a negligible CO contribution \citep{Bhardwaj02}. As 3I approaches perihelion and its water production rate rises \citep[see thermal models of][]{Puzia25, Yaginuma26}, the red lines are therefore expected to brighten more steeply than the green line. This agrees with the steeper exponents we measured for the red doublet (Fig.~\ref{fig:oxevol}, top panel). The residuals about the smooth power-law model (Fig.~\ref{fig:oxevol}, second panel) follow the same epoch-by-epoch pattern as the CN residuals. 

The G/R ratio is known to depend on nucleocentric distance because of the very different radiative lifetimes of the two red \ooned\ lines ($\tau\!\approx\!130$\,s) and the green \oones\ line \citep[$\tau\!\approx\!0.8$\,s;][]{Bhardwaj12, Raghuram14, Decock15, NIST_ASD}, which translate into very different emission scale lengths in the coma. Fiber~A of ESPRESSO covers up to a nucleocentric radius of $\sim\!650\!-\!930$\,km around the comet nucleus, and because our fiber radius is larger than the \oones\ and comparable with the \ooned\ scale length, we sampled the fully integrated emission. The variations in G/R with nucleocentric distance within the coma are therefore not resolved, and all our values are comparable among themselves. Similarly, when we compare with larger FoVs that do not cover the full \ooned\ scale length, we expect our values to be higher.

The G/R ratio in 3I decreases steadily with decreasing heliocentric distance (Fig.~\ref{fig:oxevol}, third panel), in agreement with previous cometary observations \citep{Decock13, McKay13}. All measured G/R values fall within $1\sigma$ of the piecewise model (Fig.~\ref{fig:oxevol}, fourth panel), confirming that the power-law description captures the bulk of the heliocentric trend. This behavior is consistent with thermal inertia models of the nucleus \citep{Puzia25, Yaginuma26}, in which the delayed subsurface heating causes the \coii/\hiio\ production ratio, and therefore, the G/R ratio, to decay more gradually after perihelion, owing to the higher sublimation temperature of \hiio\ relative to \coii. The comparison with a symmetric fit yields $\Delta\rm BIC$ values of $105, 2011, \text{and } 277$ for [O\,\textsc{i}]~5578, 6302, and 6365\,\AA, respectively, indicating a clear signal of asymmetry. However, the exponents of the piecewise functions are consistent within their uncertainties, and therefore, denser sampling across future cometary perihelion passages is needed to establish this asymmetry at a statistically meaningful level. Several independent studies have reported an extreme \coii/\hiio\ ratio in 3I, with \hiio\ becoming increasingly dominant as the comet approaches perihelion \citep{Lisse25, Cordiner25, Shinnaka26, Belyakov26, Lisse26}. Our G/R measurements are consistent with these findings: the ratio is elevated relative to Solar System comets at all sampled heliocentric distances (possibly due to the small aperture), and it declines steadily toward perihelion as the relative water contribution grows. Notably, our G/R values are comparable to those measured for 2I/Borisov \citep{Opitom21}, placing the two interstellar visitors in a similar region of the G/R-$r_{\rm h}$ plane, well above the locus of Jupiter-family and Oort-cloud comets.

We also note that systematic monitoring of Oort-cloud comets remains scarce, and long-term observational campaigns of such objects entering the Solar System for the first time are essential for placing future ISOs in a broader chemo-dynamical and Galactic context.

We estimated the \coii/\hiio\ ratios following the method of \citet{Decock13}, in which CO is omitted due to its low contribution, as found by the same authors. Using effective production rates from \citet{Bhardwaj12} and the branching ratio for the green line from \citet{Slanger06}, we obtained values ranging from $0.06\!\pm\!0.01$ ($r_{\rm h}\!=\!1.67$~au) to $1.06\!\pm\!0.60$ ($r_{\rm h}\!=\!2.17$~au; see Table~\ref{tab:results}). When compared directly with the G/R ratio from \citet{Shinnaka26}, our results agree, with our values slightly higher, as expected from our smaller aperture. For the \coii/\hiio\ ratios, our values agree better with the SPHEREx measurements \citep{Lisse26} and are lower than those from the JWST \citep{Belyakov26}. In particular, one of the JWST data points is inconsistent with our estimated relation (see Fig.~\ref{fig:oxevol}). This is somewhat counterintuitive: we would expect closer agreement between our values and JWST because both use smaller apertures than SPHEREx (ours was a circular aperture of radius $0.5\arcsec$, and the JWST has an annulus of $2$--$3\arcsec$). One explanation is that the conversion from the G/R ratio to \coii/\hiio\ depends on photodissociation rates that vary across the literature, which warrants further study, and they are themselves functions of heliocentric distance and potentially, solar activity. Different choices of photodissociation rates in the literature lead to a substantial spread in the derived ratios \citep[e.g.,][]{Raghuram13, McKay16, Kawakita22}. We therefore emphasize that our [O\,\textsc{i}]-based \coii/\hiio\ values are indirect proxies and that direct infrared measurements should be preferred where available. Nevertheless, the overall picture of a decreasing \coii/\hiio\ ratio with heliocentric distance and its elevated values remains consistent within the framework of our analysis. Other physical explanations for a water excess in the JWST measurements are also possible: the JWST annulus might still be insufficient to avoid the optically thick regime of \hiio, or the dominant water source might transition at these distances ($1.9-2.4$\,au) from sublimating icy grains to the nucleus surface \citep{Combi13, Bodewits14}. However, we emphasize that only one of the JWST ratios is inconsistent, which is not enough to clearly establish the cause.

As an internal consistency check, we examined the red doublet ratio $F_{6302}/F_{6365}$, whose theoretical value is set by the branching ratio of the O($^1$D)~$\to$~O($^3$P) transition, that is, 3.096 in the nonrelativistic calculation of \citet{Galavis97}, and 2.997 when relativistic corrections are included \citep{Storey00}. From the per-night line intensities listed in Table~\ref{tab:results}, nearly all measured ratios are consistent with the \citet{Storey00} value within $1\sigma$, and all but one within $2\sigma$. The sole outlier is night E3, where the 6365\,\AA\ line is severely blended with a nearby telluric feature and affected by a low S/N. The weighted mean of all nights yields $F_{6302}/F_{6365} = 2.97 \pm 0.03$.

\section{Conclusions}\label{sec:conc}
The arrival of 3I/ATLAS, the third known interstellar object to traverse the Solar System, provides a rare opportunity to probe the chemistry of another planetary system using the full arsenal of Solar System observational tools. We have monitored 3I/ATLAS with the high-resolution spectrograph ESPRESSO ($R\!\simeq\!140\,000$) on the ESO 8.2\,m VLT across both branches of its perihelion passage, with pre-perihelion observations starting in early September 2025 and post-perihelion observations until late November--December 2025, spanning $r_{\rm h}\!=\!1.67\!-\!2.45$\,au (Table~\ref{tab:observations}). The data were reduced with the standard ESPRESSO pipeline and analyzed with our publicly released Python package \textsc{CometSpec}, which we developed to perform fluorescence modeling of the CN violet band (including the Swings effect, rotational/collisional coupling, and a flexible LSF) together with an MCMC fitting, and Voigt-profile decomposition of the forbidden [O\,\textsc{i}] lines. From these fits, we derived CN production rates and the oxygen green-to-red ratio throughout the full campaign. In addition, we constructed a fluorescence model for $\rm C_2$ and Fe plus the line fitting of CH. Our main results are listed below.

\begin{itemize}

\item[$\bullet$] CN production rate evolution. We detected CN in 18 of 22 useful observed nights, with 50\% completeness upper limits on the remaining four epochs. The pre-perihelion heliocentric evolution is well described by a power law $\rm Q_{\rm CN}\!\propto\!r_{\rm h}^{\,\textrm{\it b}}$ with $b_{\rm pre}\!=\!-4.62^{+1.25}_{-1.22}$, which is within the range observed for Solar System comets \citep{AHearn95}. The post-perihelion sampling is too sparse to establish an asymmetric pre- and post-evolution at high confidence. The data are consistent with a single-branch power law and a piecewise-asymmetric model. The residual scatter about the smooth model exceeds the formal uncertainties and is plausibly driven by a combination of atmospheric, comet-centering, and Haser-model systematics plus physical effects (rotation, outbursts, and solar wind variability) that the current cadence cannot separate. The comparison with \citet{Cordiner26} is consistent with HCN being the dominant parent of CN, but other HCN measurements are inconsistent with this interpretation.

\item[$\bullet$] Forbidden oxygen lines. The three forbidden [O\,\textsc{i}] lines at 5578, 6302, and 6365\,\AA\ were detected at a high signal-to-noise ratio in nearly every night. The measured red-doublet flux ratio, $F_{6302}/F_{6365} = 2.97 \pm 0.03$, agrees with the relativistic theoretical value of \citet{Storey00}, providing an independent confirmation of our flux calibration and the absence of significant line blending (with the sole exception of epoch E3, where the 6365\,\AA\ line and a telluric line blend strongly at a low S/N).

\item[$\bullet$] G/R ratio and \coii/\hiio\ proxy. The green-to-red ratio declines steadily from $\sim\!0.42\!\pm\!0.14$ at $r_{\rm h}\!\simeq\!2.4$\,au to $\sim\!0.114\!\pm\!0.004$ near perihelion. The asymmetric piecewise model is strongly preferred over the symmetric fit ($\Delta\rm BIC\!=\!105, 2011, \text{and } 277$ for the 5578, 6302, and 6365\,\AA\ lines, respectively), with the post-perihelion branch declining more gradually than the pre-perihelion branch, as expected from the delayed \hiio\ sublimation predicted by thermal inertia models. Our G/R values are higher than in Solar System Oort-cloud and Jupiter-family comets and are comparable to those measured for 2I/Borisov, placing both interstellar visitors above the locus of Solar System comets in the G/R--$r_{\rm h}$ plane. The small aperture of our study can affect the comparison. Translated via the \citet{Decock13} prescription, the inferred \coii/\hiio\ proxy spans $\sim\!0.06$ near perihelion to $\sim\!1.06$ at $r_{\rm h} \gtrsim 2$\,au, which is consistent with the trends reported from the Subaru High Dispersion Spectrograph (Subaru/HDS), SPHEREx, and JWST, and supports a picture in which \ATLAS\ hosts an exceptionally \coii-rich coma that becomes progressively \hiio-dominated near perihelion. Denser sampling across future interstellar cometary perihelion passages is needed to establish this asymmetry at the level of individual Solar System comets.

\item[$\bullet$] Emission line inventory. Beyond CN and [O\,\textsc{i}], we identified Ni\,\textsc{i} and Fe\,\textsc{i} emission in multiple nights (Sect.~\ref{sec:lines}) and detected $\rm C_2$ Swan and CH (4300\,\AA) bands more clearly in the post-perihelion stack than in the pre-perihelion data. From the first three post-perihelion nights, we derived $\log \rm Q_{C_2}$ in the range $[25.20,\,25.68]$ and $\log \rm Q_{CH}\!=\!26.38\!\pm\!0.05$, corresponding to $\log \rm Q_{C_2}/Q_{CN} \in [-0.79,\,-0.10]$ and $\log \rm Q_{CH}/Q_{CN}\!=\!0.31\!\pm\!0.12$. These ratios place \ATLAS\ near the boundary of the carbon-chain-depleted class \citep{AHearn95}, which is broadly consistent with the pre-perihelion classifications of \citet{Rahatgaonkar25}, \citet{Salazar25}, \citet{Lazzarin26}, and \citet{Hutsemekers26}. A direct comparison of Fe\,\textsc{i} production rates between two nights at nearly identical heliocentric distances (E17 at $1.85$\,au and E21 at $1.88$\,au) yielded $\log \rm Q_{\rm Fe}\!=\!24.42\!\pm\!0.06$ and $24.41\!\pm\!0.08$, respectively, that is,~no detectable pre- and post-perihelion asymmetry. A larger matched-$r_{\rm h}$ sample, tracking iron emission from its onset to its shut off, would test the thermal models of iron-bearing carbonyl carriers \citep{Hutsemekers26} and the heliocentric thermal asymmetry inherent to such a scenario \citep{Manfroid21, Rahatgaonkar25}.

\end{itemize}

\section{Data availability}\label{sec:opensource}

To enable independent reanalysis and comparison, and to facilitate the community follow-up that the next interstellar visitor will require, \textsc{CometSpec} is publicly available on GitHub at \url{https://github.com/baltasarluco/CometSpec}; and the reduced 1D spectra (per-night and combined stacks), derived continuum models, corner plots, and fit plots are released on Zenodo at \url{https://zenodo.org/records/20966879}.

\begin{acknowledgements}
We are grateful to the ESO Garching and Paranal observatory staff for their extraordinary support during the implementation and execution of the observations, in particular we thank Martina Baratella, Susana Cerda-Hernandez, Claudia Cid, Enrico Congiu, Alex Correa,  Robert De Rosa, Julien Drevon, Lorena Faundez, Abigail Frost, Javier Fuentes, Elisa Garro, Boris Haeussler, Ana Jimenez Gallardo, Aaron Labdon, Francesca Lucertini, John Pritchard, and Jose Velasquez. We thank Dennis Bodewits for valuable discussions and comments on the manuscript, and the referee, John Noonan, for a constructive report that improved this work. We acknowledge support from the Chilean National Agency for Research and Development (ANID) in form of the grant CATA-Basal FB210003 and the Beca de Doctorado Nacional (RR, JPC).~This research has made use of data and/or services provided by the International Astronomical Union's Minor Planet Center, the NASA’s Jet Propulsion Laboratory’s Horizons System, and the NASA/IPAC Extragalactic Database, which is funded by the National Aeronautics and Space Administration and operated by the California Institute of Technology.\\ 
This work is dedicated to the memory of Vicente Maximiliano Gebauer Molina.
\end{acknowledgements}

   \bibliographystyle{aa} 
   \bibliography{aa60647-26}

\begin{appendix}
\twocolumn

\section{\textsc{CometSpec}, a flexible radiative fluorescence model package}\label{sec:CometSpec}

\textsc{CometSpec} fits the radiative fluorescence model to the spectroscopic data via MCMC using the affine-invariant ensemble sampler implemented in \textsc{emcee} \citep{Foreman13}. The user can choose which parameters are left free among the column density $\log N$ (of one or more isotopologues), the LSF parameters (e.g.,~Gaussian or Lorentzian widths), $f_{\rm col}$, and $T_{\rm kin}$. Any of these can also be held fixed.

The first 50\% of each chain is discarded as burn-in. After convergence, \textsc{CometSpec} optionally applies an iterative filter to the remaining posterior samples, retaining only those whose log-probability lies within a given threshold of the maximum log-probability reached by the sampler. The threshold starts at $5\sigma$ and is reduced at each iteration until the absolute difference between the mean and the median of the surviving chain becomes smaller than 10\% of the median value, thereby isolating the high-probability core of the posterior around the global maximum. This step can be disabled via a runtime flag.

\begin{figure*}[h!]
\centering
\includegraphics[width=1.5\columnwidth]{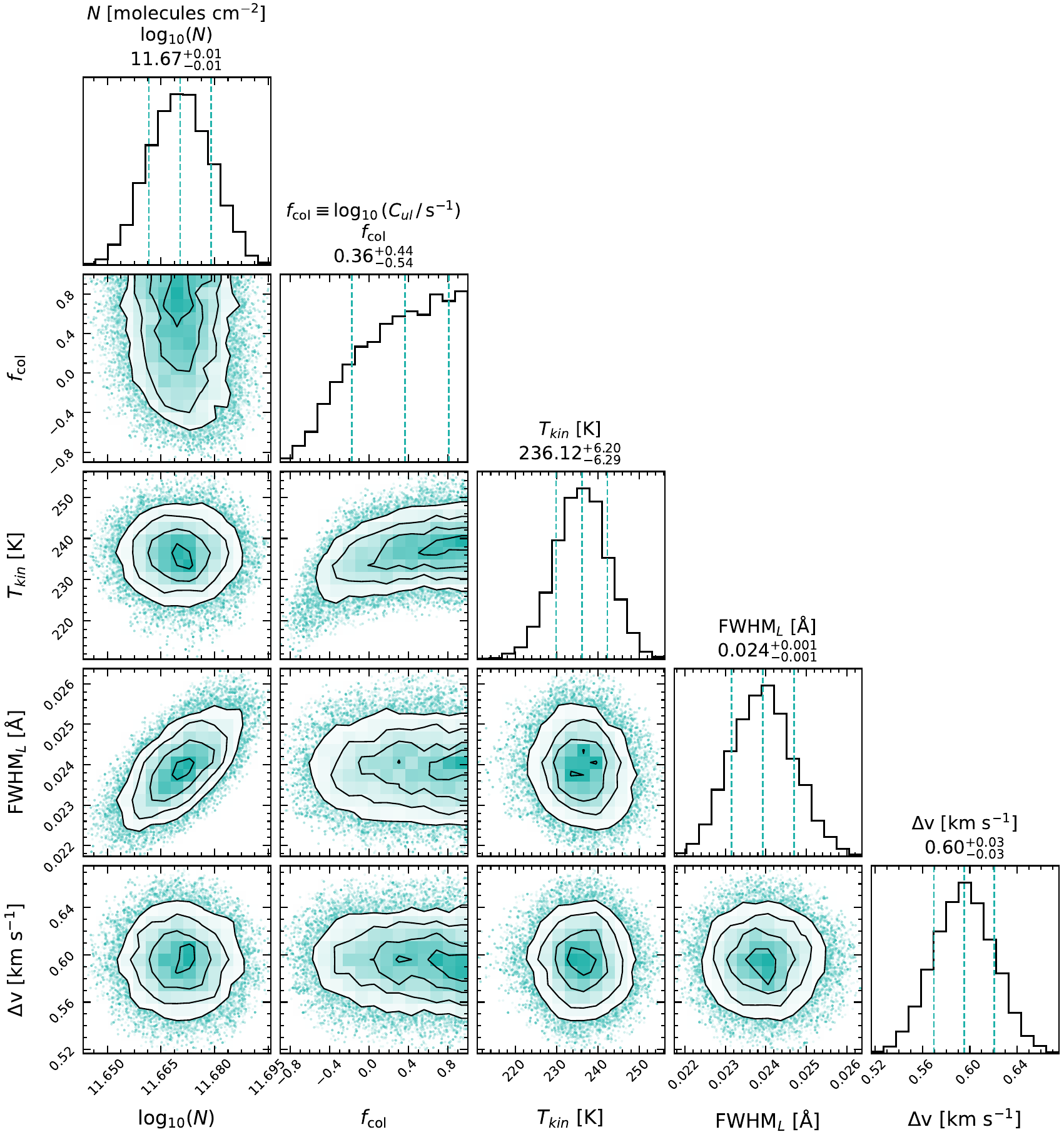}
\caption{MCMC parameter fits for the spectrum, fitting the parameters as described in Sect.~\ref{sec:cnfit}. Black contours indicate the 16th, 50th, and 84th percentiles, which are also marked as dashed cyan lines in the marginalized histograms. \label{fig:cnposterior}}
\end{figure*}

Best-fit values are reported as the median of the retained chain, and the associated uncertainties as the mean of the $|\mathrm{median}-16th|$ and $|84th-\mathrm{median}|$ percentile differences. When the iterative filter is enabled, the reported uncertainties describe the dispersion within the high-probability core and are, therefore, smaller than those derived from the full posterior. This reflects our interpretation that samples with substantially lower log-probability are treated as outliers relative to the dominant solution. The procedure yields robust best-fit values and uncertainties even when the posterior exhibits multiple converged regions, an example is shown in Fig.~\ref{fig:cnposterior}.

The iterative filter is meaningful only when the lines are clearly detected and the parameter space is well sampled around a true maximum. For non-detections, where the chain does not converge to a well-defined peak, filtering around the maximum log-probability is ill-defined, and we instead report uncertainties from the full posterior.

\section{Isotopologue search} \label{sec:isotop}

To search for isotopologues, we used the night of November 25 and the stack of the first three post-perihelion nights. We did not include additional nights in the stack, as the signal-to-noise ratio decreases owing to the exponential decline of the CN emission. Adopting the fitted $\rm ^{12}C^{14}N$ model from November 25 as a prior, we constructed models for $\rm ^{13}C^{14}N$ and $\rm ^{12}C^{15}N$ at their expected line positions. We then stacked these lines weighted by their predicted relative intensities, so that the expected stronger lines contribute more. Neither the November 25 night nor the stacked spectrum yielded a detection of any isotopologue.

\section{Fe, C$_2$ and CH}\label{app:fec2ch}

We use \textsc{CometSpec} to fit fluorescence models of Fe, $\rm C_2$, and we compute CH production rates from line fluxes.

\paragraph{Iron.} For Fe, we adopt the line list of \citet{vanHoof18}. We retrieved all transitions in the $2\,000$--$10\,000$\,\AA\ range and retained those with $A_{ul} > 10^{3}$\,s$^{-1}$ and upper-level energies below $40\,000$\,cm$^{-1}$, following \citet{Manfroid21} line filtering, with some modifications, for instance we did not apply lower-level filters and our wavelength range of lines included is different. The solar irradiance was extended into the UV using the \citet{Hall91} spectrum over $2\,000$--$2\,990$\,\AA. We then performed independent fits in successive 100\,\AA\, wide windows, with the MCMC simultaneously sampling a Lorentzian LSF width, a line velocity shift, and $\log N$ (10 walkers, $1\,000$ steps, i.e.,~$3 \times 10^{4}$ iterations per window). The Fe production rate was derived assuming a column-density profile $N(\rho) \propto \rho^{-1}$, where $\rho$ is the projected comet-centric distance \citep{Manfroid21}. Per-window uncertainties were estimated as for CN, combining statistical and slit-loss contributions and adopting the mean of the upper and lower $\log N$ errors. A single per-night $\log \rm Q_{\rm Fe}$ was then obtained as the variance-weighted mean across windows, with Gaussian error propagation. We measure $\log \rm Q_{\rm Fe}\!=\!24.42\!\pm\!0.06$ on 2025 September~24 and $24.41\!\pm\!0.08$ on 2025 December~4.

\paragraph{$\rm C_2$.} For $\rm C_2$, the default line list is the recommended ExoMol\footnote{\url{https://www.exomol.com/data/molecules/C2/}} compilation \citep{Yurchenko18, McKemmish20}. We applied the following selection criteria: wavelengths in the range $3\,000$--$10\,000$\,\AA, upper-level energies $< 30\,000$\,cm$^{-1}$, $A_{ul} > 2 \times 10^{6}$\,s$^{-1}$, vibrational quantum number $v < 5$, rotational quantum number $N < 50$, and only the $a\,^1\Pi_u - x\,^1\Sigma_g^+$, $b\,^3\Sigma_g^- - a\,^3\Pi_u$, $d\,^3\Pi_g - a\,^3\Pi_u$, $d\,^3\Pi_g - c\,^3\Sigma_u^+$, $a\,^3\Pi_u - x\,^1\Sigma_g^+$, and $c\,^3\Sigma_u^+ - x\,^1\Sigma_g^+$ transitions. The last three conditions are adopted from \citet{Rousselot00} but with more restrictive thresholds. We note that the number of transitions increases substantially if the $A_{ul}$ criterion is relaxed.

The fluorescence modeling follows the same scheme as for CN, except that the collisional selection rules depend on the isotopologue. For $^{12}\rm C_2$, only $|\Delta J| = 2$ transitions are allowed, since it is homonuclear with nuclear spin~0; for $^{13}\rm C_2$, $|\Delta J| = 0, 2$ transitions are permitted, as it is homonuclear with nuclear spin~1/2 and exhibits ortho--para separation; and for $^{12}\rm C\,^{13}C$, all $|\Delta J| = 0, 1, 2$ transitions are allowed, since it is heteronuclear. In all cases, collisions are restricted to transitions from upper levels to the lowest state in the line list (or vice versa). It is important to highlight that collisions for this model are more complicated, with significant degeneracy among the parameters.

\begin{figure}[ht!]
\centering
\includegraphics[width=\columnwidth]{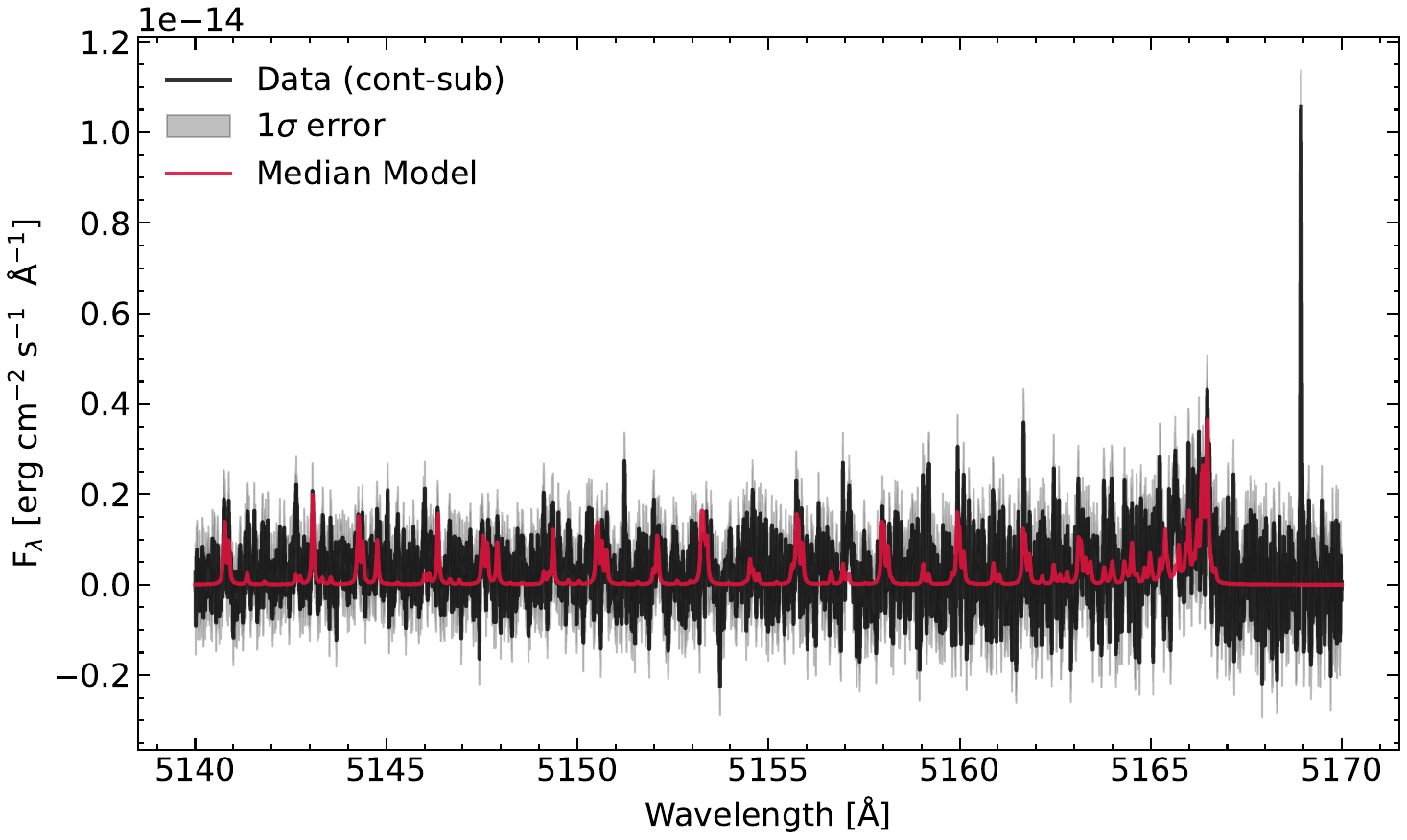}
\caption{\label{fig:c2fit}$\rm C_2$ fluorescence fit for the stacked spectrum of the first three post-perihelion nights. The black line shows the data with their 1\,$\sigma$ uncertainties; the red line is the non-collisional fluorescence fit. The prominent line at 5169\,\AA\ corresponds to a Fe line.}
\end{figure}

For $^{12}\rm C_2$ we fitted only the column density $\log N$, the LSF FWHM, and a velocity shift. The kinetic temperature was fixed at $T_{\rm kin}\!=\!100$\,K, and we ran three independent fits, with $f_{\rm col} = -3.12$, $f_{\rm col} = 8.52$, and no collisions, in order to bracket the impact of collisional thermalization, Fig.~\ref{fig:c2fit} shows the no-collision fit. These choices reflect both the high computational cost of the full model and the strong parameter degeneracies inherent to fitting a single band at low signal-to-noise. Each chain used 10 walkers and $500$ steps. We fitted the stacked spectrum of the first three post-perihelion nights, with the solar irradiance scaled to the mean heliocentric distance of those epochs. Adopting the parent scale length of $2.2\times10^4$\,km and daughter scale length of $6.6\times10^4$\,km, both scaled by $r^{-2}$ \citep{AHearn95} and the same outflow velocity as for CN, we obtain $\log \rm Q_{\rm C_2} \in [25.20,\, 25.68]$ across the three fits, corresponding to $\log (\rm Q_{\rm C_2}/Q_{\rm CN}) \in [-0.79,\, -0.10]$ when compared to the mean CN production rate over the same epochs (E18, E19 and E20). The systematic uncertainty associated with the parameter degeneracies is likely underestimated by these formal ranges.

\paragraph{CH.} For CH, the short radiative lifetime makes a full fluorescence treatment impractical, and we instead estimate the production rate from a direct flux measurement. We fitted the CH lines in the stack of the first three post-perihelion nights, summed their fluxes, converted them to column densities using Table~2 of \citet{Cochran12}, and applied a Haser model with the parent scale length of $7.8\times10^4$\,km and the daughter scale length $4.8 \times10^3$\,km  both scaled by $r^{-2}$ and from the same reference. Including the slit-loss contribution, we obtain $\log \rm Q_{\rm CH} = 26.38 \pm 0.05$ and $\log (\rm Q_{\rm CH}/Q_{\rm CN}) = 0.31 \pm 0.12$ when compared to the mean CN production rate of the same three nights (E18, E19 and E20).

\section{Continuum evolution} \label{sec:apcont}

\begin{figure*}[ht!]
\centering
\includegraphics[width=1.5\columnwidth]{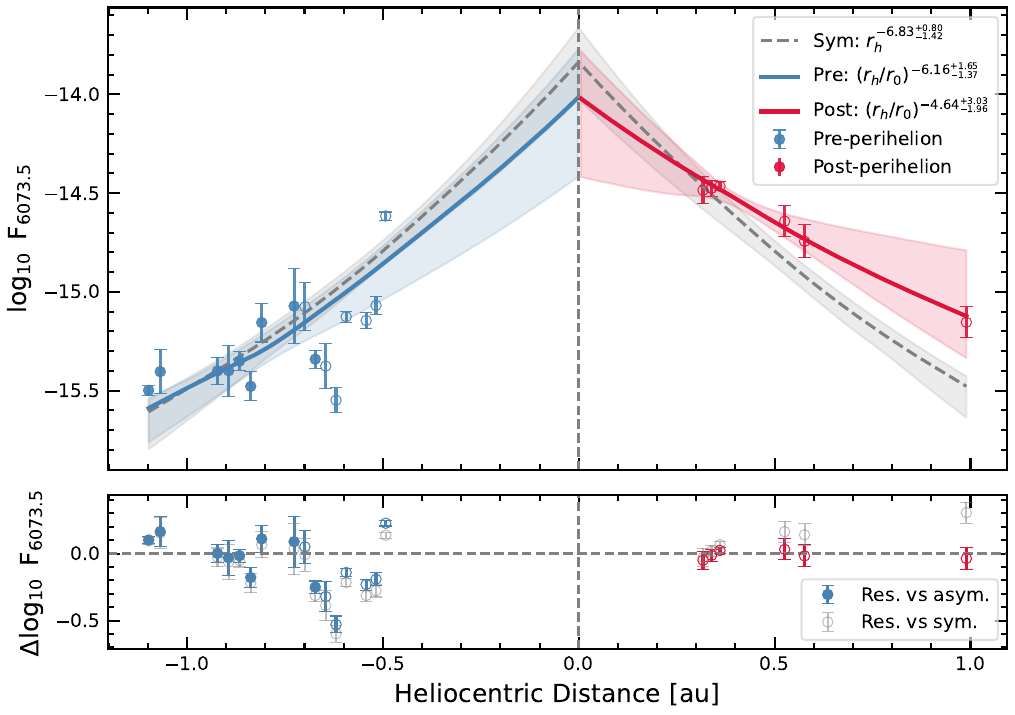}
\caption{\label{fig:continuum}Continuum evolution of 3I at different heliocentric distances. {\it Upper panel:} Continuum flux in the $6071$--$6076$\,\AA\ window (logarithmic scale) as a function of heliocentric distance $r_{\rm h}$. Blue and red symbols denote pre- and post-perihelion measurements, with $1\sigma$ error bars; the vertical dashed line marks perihelion. The black dashed line and gray band show the symmetric fit $\text{F}_{\rm 6073.5}\!=\!a\,r_{\rm h}^{\,b}$ and its $1\sigma$ confidence region; the blue and red curves show the asymmetric piecewise fit, continuous at perihelion. Open and filled symbols indicate nights without and with sky subtraction, respectively. {\it Lower panel:} Residuals with respect to the asymmetric fit (color-coded as above) and to the symmetric fit (open gray symbols); the horizontal dashed line marks zero. Colored symbol styles follow the convention of the upper panel.}
\end{figure*}

We studied the continuum evolution throughout our observations to assess the scatter present in the data. For this purpose, we computed the variance-weighted mean of the flux density in the window $6071$--$6076$\,\AA, chosen because it is relatively free of sky emission lines, telluric absorption, and solar continuum absorption. The slit-loss error was added in quadrature to the flux error, and the same fitting procedure applied to the CN evolution was then used. The resulting fit, shown in Fig.~\ref{fig:continuum}, reveals a large scatter that points toward a systematic origin. As with CN, the asymmetric profile is preferred over the symmetric one, with $\Delta \mathrm{BIC} = 60.7$, and the fit parameters are listed in Table~\ref{tab:continuum}. It is important to emphasize that the lack of sky subtraction on some nights did not affect the CN analysis, since the continuum was fitted independently; however, its contribution must be taken into account when interpreting Fig.~\ref{fig:continuum}.

\begin{table}[h]
\centering
\caption{Continuum fits of the heliocentric scaling relation $\text{F}_{\rm cont}=a\,(r_{\rm h}/1\text{ au})^{\,b}$ and $\text{F}_{\rm cont} = \text{F}_{\rm peri}\,(r_{\rm h}/r_0)^{\,b}$}
\label{tab:continuum}
\setlength{\extrarowheight}{2pt}
\begin{tabular}{cccc}
\hline\hline
Data included & $\log a$ & $b$ & $\log \mathrm{F_{peri}}$ \\
\hline
Both & $-12.93^{+0.36}_{-0.29}$ & $-6.83^{+0.80}_{-1.42}$ & --- \\
Pre  & --- & $-6.16^{+1.65}_{-1.37}$ & $-14.01^{+0.24}_{-0.40}$ \\
Post & --- & $-4.64^{+3.03}_{-1.96}$ & $-14.01^{+0.24}_{-0.40}$ \vspace{0.5mm}\\
\hline\hline\\
\end{tabular}
\vspace{-1ex}
\begin{minipage}{\textwidth}
\end{minipage}
\end{table}

\onecolumn

\section{Tables}\FloatBarrier
\begin{table*}[h]
\centering
\caption{Journal of 3I observations with VLT/ESPRESSO.}
\label{tab:observations}
\begin{tabular}{l l l r r r r r r r }
\hline\hline
Visit \tablefootmark{a}  & Date & Target       & Exp. Time & Air mass \tablefootmark{b} & Seeing \tablefootmark{c} & $r_\mathrm{h}$ \tablefootmark{d} & $\dot{r}$\tablefootmark{d} & $\Delta$ \tablefootmark{e}& $\dot{\Delta}$ \tablefootmark{e} \\

& (UT) & & (s) &  & (\arcsec) & (au) & (km\,s$^{-1}$) & (au) & (km\,s$^{-1}$)\\
\hline
$\star$E1-1  & 2025-09-02 23:37--02:19 & \ATLAS     & $9\,\times\,600$  & $1.26-2.08$ & $0.40-0.53$ & $2.45$ & $-51.56$ & $2.57$ & $-4.51$  \\
    &                         & Sky        & $2\,\times\,600$  & --- & --- & --- & --- & --- & --- \\
    &                         & \ObjSolarA & $1\,\times\,600$  & --- & --- & --- & --- & --- & --- \\

$\star$E2-1  & 2025-09-03 23:24--02:04 & \ATLAS     & $12\,\times\,600$ & $1.22-2.64$ & $0.44-0.82$ & $2.42$ & $-51.30$ & $2.57$ & $-4.28$ \\
    &                         & Sky        & $2\,\times\,300$  & --- & --- & --- & --- & --- & --- \\

$\star$E3-2  & 2025-09-08 00:26--01:13 & \ATLAS     & $3\,\times\,600$  & $1.81-2.09$ & $0.98-1.12$ & $2.28$ & $-49.85$ & $2.55$ & $-3.47$ \\
    &                         & Sky        & $1\,\times\,300$  & --- & --- & --- & --- & --- & --- \\

$\star$E4-1 \tablefootmark{f}  & 2025-09-09 23:34--01:17 & \ATLAS     & $7\,\times\,600$  & $1.48-2.45$ & $0.92-1.10$ & $2.25$ & $-49.53$ & $2.55$ & $-3.40$ \\
    &                         & Sky        & $1\,\times\,300$  & --- & --- & --- & --- & --- & --- \\

$\star$E5-1  & 2025-09-10 23:33--00:22 & \ATLAS     & $3\,\times\,600$  & $1.50-1.66$ & $0.74-0.88$ & $2.22$ & $-49.20$ & $2.55$ & $-3.35$ \\
    &                         & Sky        & $1\,\times\,300$  & --- & --- & --- & --- & --- & --- \\

$\star$E6-1  & 2025-09-11 23:27--01:01 & \ATLAS     & $6\,\times\,600$  & $1.49-2.19$ & $0.68-0.82$ & $2.19$ & $-48.84$ & $2.55$ & $-3.28$ \\
    &                         & Sky        & $2\,\times\,300$  & --- & --- & --- & --- & --- & --- \\

$\star$E7-2  & 2025-09-12 23:41--00:33 & \ATLAS     & $3\,\times\,600$  & $1.66-1.88$ & $0.60-0.99$ & $2.17$ & $-48.47$ & $2.55$ & $-3.24$ \\
    &                         & Sky        & $1\,\times\,300$  & --- & --- & --- & --- & --- & --- \\

$\star$E8-2  & 2025-09-15 23:21--00:26 & \ATLAS     & $4\,\times\,600$  & $1.64-2.04$ & $0.78-1.40$ & $2.08$ & $-47.25$ & $2.54$ & $-3.27$  \\
    &                         & Sky        & $1\,\times\,300$  & --- & --- & --- & --- & --- & --- \\

E9-2  & 2025-09-16 23:49--00:42 & \ATLAS     & $4\,\times\,600$  & $1.98-2.64$ & $0.94-1.18$ & $2.06$ & $-46.81$ & $2.54$ & $-3.30$ \\

$\star$E10-1 & 2025-09-17 23:26--00:28 & \ATLAS     & $4\,\times\,600$  & $1.75-2.26$ & $0.74-0.78$ & $2.03$ & $-46.36$ & $2.54$ & $-3.38$ \\
    &                         & Sky        & $4\,\times\,300$  & --- & --- & --- & --- & --- & --- \\

E11-1 & 2025-09-18 23:11--00:33 & \ATLAS     & $6\,\times\,600$  & $1.74-2.74$ & $0.71-0.90$ & $2.00$ & $-45.88$ & $2.53$ & $-3.46$ \\

E12-1 & 2025-09-19 23:35--00:11 & \ATLAS     & $3\,\times\,600$  & $2.01-2.37$ & $0.86-0.94$ & $1.98$ & $-45.38$ & $2.53$ & $-3.57$ \\

E13-1 & 2025-09-20 23:21--23:47 & \ATLAS     & $3\,\times\,600$  & $1.92-2.29$ & $0.58-0.64$ & $1.95$ & $-44.86$ & $2.53$ & $-3.70$ \\

E14-1 \tablefootmark{g} & 2025-09-21 23:02--23:48 & \ATLAS     & $2\,\times\,600$  & $1.88-2.24$ & $0.71-0.89$ & $1.93$ & $-44.32$ & $2.53$ & $-3.84$ \\

E15-1 & 2025-09-22 23:17--23:53 & \ATLAS     & $3\,\times\,600$  & $2.02-2.37$ & $0.63-0.72$ & $1.90$ & $-43.75$ & $2.52$ & $-4.00$ \\

E16-1 & 2025-09-23 23:22--23:58 & \ATLAS     & $3\,\times\,600$  & $2.18-2.62$ & $0.66-0.75$ & $1.87$ & $-43.15$ & $2.52$ & $-4.17$ \\

E17-2 & 2025-09-24 23:10--23:56 & \ATLAS     & $3\,\times\,600$  & $2.26-2.73$ & $0.48-0.53$ & $1.85$ & $-42.53$ & $2.52$ & $-4.37$ \\

E18-2 & 2025-11-25 08:20--08:56 & \ATLAS     & $3\,\times\,600$  & $2.07-2.44$ & $0.63-0.79$ & $1.67$ & $36.81$ & $1.98$ & $-21.67$ \\

E19-3 & 2025-11-26 08:21--09:02 & \ATLAS     & $3\,\times\,600$  & $1.91-2.21$ & $0.40-0.64$ & $1.70$ & $37.67$ & $1.97$ & $-21.33$  \\

E20-2 & 2025-11-27 08:26--09:03 & \ATLAS     & $6\,\times\,600$  & $1.82-2.08$ & $0.55-0.62$ & $1.72$ & $38.48$ & $1.95$ & $-20.96$ \\

E21-3 & 2025-12-04 07:22--08:35 & \ATLAS     & $6\,\times\,600$  & $1.64-2.35$ & $0.64-0.80$ & $1.88$ & $43.35$ & $1.88$ & $-17.03$\\

E22-3 \tablefootmark{h} & 2025-12-06 07:29--08:26 & \ATLAS     & $6\,\times\,600$ & $1.61-2.10$ & $0.78-0.95$ & $1.93$ & $44.50$ & $1.86$ & $-15.44$ \\
E23-1 & 2025-12-21 05:47--07:18 & \ATLAS     & $7\,\times\,600$  & $1.41-1.87$ & $0.54-0.86$ & $2.35$& $50.56$ & $1.80$ & $2.82$ \\ 
\hline\hline\\
\end{tabular}

\tablefoot{
\tablefoottext{a}{Observing visit identifiers for ESPRESSO, E + epoch number - UT telescope number, nights with sky subtraction start with a $\star$ symbol. Each entry corresponds to a distinct observing night, including comet and if applicable solar analog and/or sky exposures.}
\tablefoottext{b}{Air mass range of the target during the science exposures.}
\tablefoottext{c}{DIMM seeing values from Paranal ambient telemetry during the science exposures.}
\tablefoottext{d}{Mean heliocentric distance and velocity of the target during science exposures (JPL Horizons).}
\tablefoottext{e}{Mean geocentric distance and velocity of the target during science exposures (JPL Horizons).}
\tablefoottext{f}{Due to star contamination, one exposure was discarded.}
\tablefoottext{g}{Due to technical problems, compromised-quality data were taken during the night.}
\tablefoottext{h}{Due to bad weather conditions the first exposure was discarded.} 
}
\end{table*}

\begin{sidewaystable*}
\centering
\caption{3I Fit results}
\label{tab:results}

\begin{tabular}{l c c c c c c c c c c}
\hline\hline
Visit  & $\log \rm Q_{\rm CN} /(molecules \,s^{-1})$ & \multicolumn{3}{c}{Oxygen Flux (10$^{-16}$ erg cm$^{-2}$ s$^{-1}$)} & G/R & $\rm CO_2/\rm H_2O$\,\tablefootmark{a} & \multicolumn{3}{c}{Fitted $\dot{\Delta}$ (km s$^{-1}$)} \\
 & & 5578\,\AA & 6302\,\AA & 6365\,\AA & & & 5578\,\AA & 6302\,\AA & 6365\,\AA \\
\hline
E1  & $\leq25.39$ & $1.21\pm0.40$ & $2.06\pm0.21$ & $0.79\pm0.33$ & $0.42\pm0.14$ & -- & $-3.84\pm0.63$ & $-4.53\pm0.13$ & $-4.22\pm0.85$\\
E2  & $\leq25.55$ & $1.67\pm0.24$ & $2.87\pm0.23$ & $0.90\pm0.33$ & $0.43\pm0.08$ & -- & $-3.90\pm0.24$ & $-4.47\pm0.11$ & $-4.40\pm0.72$ \\
E3  & $\leq 25.95$& $3.42\pm0.76$ & $4.14\pm0.52$ & $3.30\pm1.22$ & $0.45\pm0.12$ & -- & $-3.29\pm0.57$ & $-3.84\pm0.15$ & $-2.89\pm0.89$ \\
E4  & $25.61\pm 0.18$ & $2.48\pm0.38$ & $4.43\pm0.48$ & $1.30\pm0.43$ & $0.42\pm0.08$ & -- & $-3.47\pm0.17$ & $-3.68\pm0.18$ & $-3.58\pm0.60$ \\
E5  & $25.61\pm 0.08$ & $2.31\pm0.43$ & $5.15\pm0.74$ & $1.46\pm0.48$ & $0.34\pm0.08$ & $0.88\pm0.60$ & $-3.54\pm0.19$ & $-3.60\pm0.18$ & $-3.80\pm0.43$ \\
E6  & $25.51\pm 0.14$ & $2.08\pm0.23$ & $5.10\pm0.50$ & $1.91\pm0.59$ & $0.29\pm0.04$ & $0.57\pm0.22$ & $-3.51\pm0.10$ & $-3.58\pm0.14$ & $-3.14\pm0.57$ \\
E7  & $25.79\pm 0.18$ & $4.99\pm0.69$ & $9.94\pm0.94$ & $3.60\pm0.83$ & $0.36\pm0.06$ & $1.06\pm0.60$ & $-3.47\pm0.22$ & $-3.36\pm0.14$ & $-3.34\pm0.38$ \\
E8  & $26.07\pm 0.26$ & $5.18\pm0.60$ & $17.63\pm1.09$ & $5.52\pm1.24$ & $0.22\pm0.03$ & $0.31\pm0.09$ & $-3.59\pm0.19$ & $-3.27\pm0.10$ & $-3.13\pm0.37$ \\
E9  & $26.15\pm 0.15$ & $6.47\pm0.66$ & $15.69\pm0.71$ & $5.37\pm0.72$ & $0.31\pm0.03$ & $0.65\pm0.19$ & $-3.35\pm0.09$ & $-3.64\pm0.06$ & $-3.56\pm0.16$ \\
E10 & $25.80\pm 0.07$ & $3.19\pm0.35$ & $9.61\pm0.66$ & $2.70\pm0.51$ & $0.26\pm0.03$ & $0.43\pm0.13$ & $-3.60\pm0.16$ & $-3.45\pm0.10$ & $-3.48\pm0.20$ \\
E11 & $25.78\pm 0.18$ & $3.49\pm0.42$ & $9.96\pm0.49$ & $3.92\pm0.65$ & $0.25\pm0.03$ & $0.41\pm0.12$ & $-3.60\pm0.19$ & $-3.58\pm0.08$ & $-3.25\pm0.33$ \\
E12 & $25.56\pm 0.07$ & $1.86\pm0.42$ & $6.63\pm0.60$ & $1.85\pm0.51$ & $0.22\pm0.05$ & $0.29\pm0.15$ & $-3.95\pm0.28$ & $-3.71\pm0.13$ & $-3.75\pm0.34$ \\
E13 & $25.89\pm 0.06$ & $4.49\pm0.52$ & $16.60\pm0.67$ & $5.18\pm0.67$ & $0.21\pm0.03$ & $0.26\pm0.07$ & $-3.93\pm0.20$ & $-3.61\pm0.07$ & $-3.63\pm0.20$ \\
E15 & $25.69\pm 0.07$ & $4.04\pm0.34$ & $12.52\pm0.41$ & $4.30\pm0.55$ & $0.24\pm0.02$ & $0.36\pm0.08$ & $-4.05\pm0.15$ & $-3.97\pm0.05$ & $-3.82\pm0.20$ \\
E16 & $25.90\pm 0.08$ & $4.65\pm0.44$ & $18.01\pm0.45$ & $5.54\pm0.34$ & $0.20\pm0.02$ & $0.24\pm0.05$ & $-4.16\pm0.13$ & $-4.14\pm0.04$ & $-4.21\pm0.08$ \\
E17 & $26.19\pm 0.06$ & $13.23\pm0.64$ & $45.54\pm0.60$ & $14.94\pm0.45$ & $0.22\pm0.01$ & $0.30\pm0.03$ & $-4.33\pm0.08$ & $-4.31\pm0.02$ & $-4.30\pm0.04$ \\
E18 & $25.95\pm 0.11$ & $10.45\pm0.34$ & $68.58\pm0.69$ & $23.23\pm0.46$ & $0.114\pm0.004$ & $0.06\pm0.01$ & $-21.85\pm0.06$ & $-21.64\pm0.02$ & $-21.66\pm0.05$ \\
E19 & $25.79\pm 0.14$ & $9.85\pm0.32$ & $54.79\pm0.47$ & $18.88\pm0.41$ & $0.134\pm0.004$ & $0.10\pm0.01$ & $-21.44\pm0.09$ & $-21.30\pm0.02$ & $-21.37\pm0.06$ \\
E20 & $25.91\pm 0.06$ & $9.54\pm0.43$ & $52.25\pm0.52$ & $18.05\pm0.49$ & $0.136\pm0.006$ & $0.10\pm0.01$ & $-20.95\pm0.10$ & $-20.95\pm0.02$ & $-20.87\pm0.06$ \\
E21 & $25.83\pm 0.14$ & $10.45\pm0.28$ & $43.26\pm0.37$ & $13.91\pm0.30$ & $0.183\pm0.005$ & $0.20\pm0.01$ & $-17.13\pm0.06$ & $-17.03\pm0.03$ & $-17.01\pm0.07$ \\
E22 & $25.76\pm 0.13$ & $5.30\pm0.30$ & $25.97\pm0.34$ & $9.25\pm0.29$ & $0.150\pm0.009$ & $0.13\pm0.02$ & $-15.51\pm0.12$ & $-15.54\pm0.02$ & $-15.47\pm0.06$ \\
E23 & $\leq 25.53$& $2.05\pm0.53$ & $6.41\pm1.29$ & $1.76\pm0.57$ & $0.25\pm0.08$ & $0.40\pm0.28$ & $2.86\pm0.47$ & $2.61\pm0.36$ & $2.79\pm0.54$ \\
\hline\hline\\
\end{tabular}

\tablefoot{
\tablefoottext{a}{Using the equations from \cite{Decock13} and error propagation, the first 4 rows are omitted as they are closer to the asymptotic value.
}}
\end{sidewaystable*}

\end{appendix}
\end{document}